%% file: sunburst_Sharon22_arXiv.tex
\documentclass[twocolumn]{aastex63}

\usepackage{amsmath}
\usepackage{natbib}
\usepackage{color,soul}
\usepackage{url}
\usepackage{lineno}

\definecolor{LightCyan}{rgb}{0.88,1,1}

\newcommand{\lenstool}{{\tt{Lenstool}}}

\newcommand{\Sunburstarc}{Sunburst Arc}

\newcommand{\hst}{{\it HST}}
\newcommand{\HST}{{\it HST}}
\newcommand{\JWST}{{\it JWST}}
\newcommand{\HSTlong}{{\it Hubble Space Telescope}}

\newcommand{\planck}{{\it Planck}}

\newcommand{\chandra}{{\it Chandra}}
\newcommand{\Chandra}{{\it Chandra}}
\newcommand{\vlt}{{Very Large Telescope}}
\newcommand{\muse}{{MUSE}}
\newcommand{\MUSE}{{MUSE}}
\newcommand{\MUSElong}{{Multi Unit Spectroscopic Explorer}}
\newcommand{\magelong}{{Magellan Echellette}}
\newcommand{\firelong}{{Folded-port InfraRed Echelle}}

\newcommand{\zspec}{z_{spec}}

\newcommand{\Lya}{Ly$\alpha$}
\newcommand{\lya}{Ly$\alpha$}

\newcommand{\OIII}{[O III]~4959, 5007\,\AA}

\newcommand{\HeI}{\hbox{{\rm He}\kern 0.1em{\sc i}}}
\newcommand{\MgI}{\hbox{{\rm Mg}\kern 0.1em{\sc i}}}
\newcommand{\MgII}{\hbox{{\rm Mg}\kern 0.1em{\sc ii}}}
\newcommand{\FeII}{\hbox{{\rm Fe}\kern 0.1em{\sc ii}}}

\newcommand{\msun}{M$_{\odot}$}

\newcommand{\northwest}{northwest}
\newcommand{\northeast}{northeast}
\newcommand{\southwest}{west}
\newcommand{\west}{west}
\newcommand{\north}{north}
\newcommand{\southeast}{southeast}

\newcommand{\clustername}{PSZ1\,G311.65$-$18.48}
\newcommand{\zcluster}{0.443}
\newcommand{\zarc}{2.3703} 
\newcommand{\Ngal}{250}
\newcommand{\clustermass}{$2.93 ^{+0.01}_{-0.02} \times 10^{14}$\msun}

\received{July 2, 2022}
\revised{September 7, 2022}
\submitjournal{ApJ}

\shorttitle{Lensing Analysis of PSZ1\,G311.65$-$18.48 and the Sunburst Arc}
\shortauthors{Sharon et al.}

\graphicspath{{./}{figures/}}

\begin{document}

\title{The Cosmic Telescope that Lenses the Sunburst Arc, PSZ1\,G311.65$-$18.48: Strong Gravitational Lensing model and Source Plane Analysis\footnote{Based on observations made with the NASA/ESA {\it Hubble Space Telescope}, obtained at the Space Telescope Science Institute, which is operated by the Association of Universities for Research in Astronomy, Inc., under NASA contract NAS 5-26555. These observations are associated with programs GO-15101, GO-15377, GO-15418, GO-15949} }

\correspondingauthor{Keren Sharon}
\email{kerens@umich.edu}

\author[0000-0002-7559-0864]{Keren Sharon}
\affiliation{Department of Astronomy, University of Michigan, 1085 S. University Ave, Ann Arbor, MI 48109, USA}

\author[0000-0003-3266-2001]{Guillaume Mahler}
\affiliation{Department of Astronomy, University of Michigan, 1085 S. University Ave, Ann Arbor, MI 48109, USA}

\author[0000-0002-9204-3256]{T. Emil Rivera-Thorsen}
\affiliation{The Oskar Klein Centre, Department of Astronomy, Stockholm University, AlbaNova, SE-10691 Stockholm, Sweden}

\author[0000-0003-2200-5606]{H{\aa}kon Dahle}
\affiliation{Institute of Theoretical Astrophysics, University of  Oslo,  P. O. Box 1029, Blindern, N-0315 Oslo, Norway }

\author[0000-0003-1370-5010]{Michael D. Gladders}
\affiliation{Department of Astronomy and Astrophysics, University of Chicago, 5640 South Ellis Avenue, Chicago, IL 60637, USA}
\affiliation{Kavli Institute for Cosmological Physics, University of Chicago, 5640 South Ellis Avenue, Chicago, IL 60637, USA}

\author[0000-0003-1074-4807]{Matthew B. Bayliss}
\affiliation{Department of Physics, University of Cincinnati, Cincinnati, OH 45221, USA}

\author[0000-0001-5097-6755]{Michael K. Florian}
\affiliation{Steward Observatory, University of Arizona, 933 North Cherry Ave., Tucson, AZ 85721, USA}

\author[0000-0001-6505-0293]{Keunho J. Kim}
\affiliation{Department of Physics, University of Cincinnati, Cincinnati, OH 45221, USA}

\author[0000-0002-3475-7648]{Gourav Khullar}
\affiliation{Department of Astronomy and Astrophysics, University of Chicago, 5640 South Ellis Avenue, Chicago, IL 60637, USA}
\affiliation{Kavli Institute for Cosmological Physics, University of Chicago, 5640 South Ellis Avenue, Chicago, IL 60637, USA}
\affiliation{Kavli Institute for Astrophysics \& Space Research, Massachusetts Institute of Technology, 77 Massachusetts Ave., Cambridge, MA 02139, USA}

\author[0000-0003-0094-6827]{Ramesh Mainali}
\affiliation{Observational Cosmology Lab, Code 665, NASA Goddard Space Flight Center, 8800 Greenbelt Rd., Greenbelt, MD 20771, USA}

\author[0000-0003-4470-1696]{Kate A. Napier}
\affiliation{Department of Astronomy, University of Michigan, 1085 S. University Ave, Ann Arbor, MI 48109, USA}
\author[0000-0001-7548-0473]{Alexander Navarre}
\affiliation{Department of Physics, University of Cincinnati, Cincinnati, OH 45221, USA}
\author[0000-0002-7627-6551]{Jane R. Rigby}
\affiliation{Observational Cosmology Lab, Code 665, NASA Goddard Space Flight Center, 8800 Greenbelt Rd., Greenbelt, MD 20771, USA}

\author[0000-0002-7868-9827]{Juan David Remolina Gonz\'{a}lez}
\affiliation{Department of Astronomy, University of Michigan, 1085 S. University Ave, Ann Arbor, MI 48109, USA}

\author[0000-0001-9851-8753]{Soniya Sharma}
\affiliation{Observational Cosmology Lab, Code 665, NASA Goddard Space Flight Center, 8800 Greenbelt Rd., Greenbelt, MD 20771, USA}

\begin{abstract}
We present a strong lensing analysis of the cluster \clustername, based on \HSTlong\ imaging, archival VLT/MUSE spectroscopy, and \chandra\ X-ray data.  This cool-core  cluster ($z=\zcluster$) lenses the brightest lensed galaxy known, dubbed the ``\Sunburstarc'' ($z=2.3703$), a  Lyman continuum (LyC) emitting galaxy multiply-imaged 12 times. We identify in this field 14 additional strongly-lensed galaxies to constrain a strong lens model, and report secure spectroscopic redshifts of four. We measure a projected cluster core mass of $M(<250 {\rm kpc})=$ \clustermass. The two least-magnified but complete images of the Sunburst Arc’s  source galaxy are magnified by $\sim13\times$, while the LyC clump is magnified by $\sim$4-80$\times$. We present time delay predictions and conclusive evidence that a discrepant clump in the Sunburst Arc, previously claimed to be a transient, is not variable, thus strengthening the hypothesis that it results from an exceptionally high magnification. A source plane reconstruction and analysis of the Sunburst Arc finds its physical size to be $1\times2$kpc, and that it is resolved in three distinct directions in the source plane, $0^\circ$, $40^\circ$, and $75^\circ$ (east of North). We place an upper limit of $r\lesssim50$pc on the source plane size of unresolved clumps, and $r\lesssim 32$pc for the LyC clump. Finally, we report that the Sunburst Arc is likely in a system of two or more galaxies separated by $\lesssim6$kpc in projection. Their interaction may drive star formation and could play a role in the mechanism responsible for the leaking LyC radiation.

\end{abstract}

\keywords{Gravitational lensing: strong --- Galaxies: clusters: individual:\clustername}

\section{Introduction} \label{sec:intro}

Nature's most powerful telescopes, strong gravitational lenses, are now routinely used as a tool to study the highly-magnified universe behind them. Observations of gravitationally-lensed galaxies at $z\sim 1-3$ can probe spatial scales on the order of tens of parsecs \citep[e.g.,][]{johnson17,james18}, and can enable spectral diagnostics \citep[e.g.,][]{rigby2018,chisholm19,fischer19,patricio19}  that are otherwise beyond the capabilities of current observatories. 

Since gravitational lensing is wavelength independent, the interpretation of many observed physical properties of lensed sources is lensing invariant. These include all measurements that rely on color, line ratios, optical depth, and wavelength --- like stellar population ages, metallicity, and velocities. In addition, in calculating some relative quantities such as specific star formation rate, the lensing magnification cancels out, making the quotient independent of the magnification. 

By contrast, the interpretation of absolute measurements --- like luminosity, star formation rate, and stellar mass --- relies heavily on properly understanding how gravitational lensing affects the observed quantities, as they are sensitive to the details of the transformation of measured properties from the observed frame to the intrinsic frame of the source. 

The strong lensing galaxy cluster \clustername, which is the focus of this paper, was discovered by \citet{dahle16}, in an optical imaging program to follow up the \planck\ catalog of Sunyaev Zel'dovich (SZ) cluster candidates. The imaging data confirmed that the candidate is indeed a cluster, and revealed a $55\farcs0$-long giant arc, which is projected $25''-34''$ from the brightest cluster galaxy (BCG), and with an azimuthal extent of $108 ^{\circ}$. 
\citet{dahle16} reported integrated Vega magnitudes of R, z, J, Ks = 17.82, 17.38, 16.75, 15.43 mag, making it brighter by more than one magnitude than any previously known lensed galaxy at $z=2-3$. 
Shallow long-slit spectroscopy of the giant arc, using the Inamori-Magellan Areal Camera \& Spectrograph (IMACS) instrument on the Magellan-I 6.5-m telescope (spectral resolution $R\simeq700$), revealed numerous emission and absorption lines, including \Lya\ in emission, Si~II~$\lambda\lambda1259,1294$ \,\AA, C~IV $\lambda\lambda~1548,1550$\,\AA, and nebular semi-forbidden transitions of O III] $\lambda1666$\,\AA, Si III] $\lambda 1892$\,\AA, and C III] $\lambda\lambda1907, 1909$\,\AA\ at $z= 2.3686 \pm 0.0006$. Observations targeting the foreground cluster lens include a spectrum of the BCG, which places it at redshift $z=0.44316 \pm 0.00035$. The BCG spectrum interestingly reveals strong emission of the [OII] $\lambda\lambda3727,3729$\,\AA\ doublet, which is generally an indication of dust-unobscured star formation activity \citep{Calzetti2012}.
\citet{dahle16} estimated the mass of the core of the cluster from the projected radial distance of the giant arc, assuming spherical symmetry. The total projected mass density enclosed within the radius of the arc was estimated as $M = 1.8\pm 0.6 \times 10^{14}$\msun, which, with crude extrapolation of the cluster core mass to larger radii, is consistent with the estimated SZ-inferred mass.

\citet{rivera-thorsen17} observed the brightest region of the arc with the \magelong\ (MagE) and with the \firelong\ (FIRE) instruments on the Magellan-I 6.5-m telescope, obtaining broad wavelength coverage and higher spectral resolution than the previous IMACS data. They found that the \lya\ emission is triple-peaked, which they interpreted as a combination of emission from \lya\ that underwent a large number of scatterings, and a narrow component that emerges directly from the source without scattering. To explain this triple-peaked profile, they proposed that the galaxy has a perforated neutral medium, where some radiation escapes the source directly through an ionized channel, while some radiation experiences multiple scatterings in an optically thick neutral medium. They nicknamed the source ``the \Sunburstarc'', for its resemblance to a ``direct view of the Sun through rifted clouds''.

In a follow-up work, using data from the \HSTlong\ (\hst),  \citet{rivera-thorsen19} found that large quantities of ionizing photons are escaping from this lensed galaxy, by targeting the field with the \hst/WFC3 F275W broad-band filter (GO-15418).  At the redshift of the giant arc, the  F275W  filter captures the $\lambda < 912_{\rm rest}$~\AA\ photons that are capable of ionizing hydrogen --- also called the Lyman Continuum (LyC).  Given the compact region exhibiting LyC escape, and the \lya\ profile, 
\citet{rivera-thorsen19} concluded that the observed emission is consistent with LyC radiation escaping through a narrow clear channel in otherwise optically thick gas.
The leaked LyC radiation is coming from a star forming region within the source galaxy, which is strongly lensed into 12 images. The images of the LyC source are unresolved in the \hst\ data, and appear as point sources. The  variation of up to a factor of five in the apparent escape fraction measured in the different copies of the source was attributed to varying absorption by neutral hydrogen along the different lines of sight to these images, thus probing the patchiness of the intergalactic or circumgalactic medium.

\citet{chisholm19} fit \texttt{Starburst99} models to MagE spectroscopy (see \autoref{sec:spectroscopy}) of the LyC knot, and derived its UV-light-weighted age and stellar metallicity, finding an age of 3~Myr and $Z_*=0.55\pm0.04 \rm{Z}_{\odot}$.

\citet{vanzella20a} used lensing-symmetry arguments to estimate a model-independent lower limit of the average magnification of the arc, $\mu>20$. 
They presented archival \vlt\ (VLT) \MUSElong\ (\muse) spectroscopy (see \autoref{sec:spectroscopy}), and, absent a lens model, derived the physical properties of the source as a function of the unknown magnification, based on the age and metallicity measured by \citet{chisholm19}. They concluded that the LyC knot is a gravitationally bound star cluster, with an effective radius smaller than 20 pc, and stellar mass in the range $10^6-10^7$\msun\ depending on magnification and initial mass function, implying that it is very massive and very dense, and that the LyC leaking radiation is highly localized.

Using the $z=\zarc$ giant arc as an extended backlight, \citet{lopez20} studied the  absorbing circumgalactic medium halo of a foreground galaxy at $z=0.7$. The interloper is seen in the \hst\ imaging directly, and its gas halo is revealed in the \MUSE\ and MagE  spectra of the giant arc as absorption lines from \MgII, \FeII, and \MgI. The background arc, which probes the gaseous envelope  of the intervening absorber from $0-30$ kpc, enables a dense sampling of the rotation curve, the gas distribution, and its enrichment profile. This study adds to a growing literature of using extended arcs as backlight, which complements the traditional pencil-beam approach of using background quasars as a light source to study intervening absorbers. The absorber plane impact parameters, i.e., distances between segments of the \Sunburstarc\ images and the studied foreground galaxy in \citet{lopez20} were measured using an early version of the lens model presented here.

{Two lens models have been published for this system in the past year. \citet{pignataro21} used a subset of the \hst\ data we present here, and the same archival \MUSE\ data, to identify lensing constraints and measure spectroscopic redshifts of lensed galaxies. Their parametric lens model is based on constraints from four multiply-imaged lensed galaxies. They also used the velocity dispersion of several cluster-member galaxies to inform their contribution to the lensing potential. They estimated a cluster core mass of $\sim 2\times 10^{14}$\msun\ within $\sim200$~kpc, and found that the mass distribution is fairly symmetrical, with contribution of less than 10\% from subhalos. }

Building on the lensing evidence identified by \citet{pignataro21}, \citet{diego22} modeled the cluster with the hybrid algorithm WSLAP$+$ \citep{diego2005}. They added strong lensing constraints on the positions of the critical curve as identified from the observed symmetry in the giant arcs. Their analysis pays close attention to the non-trivial lensing configuration in the \northwest\ image of the \Sunburstarc\ in \clustername, and in particular explores models that can produce the high multiplicity and the appearance of a discrepant point source by forcing the critical curve to pass through certain positions. They reported limits on the mass and location of the substructure that is required in order to explain the observed magnifications and morphology within this arc. \citet{diego22} also presented time delay and magnification predictions in their analysis and concluded that the discrepant clump is unlikely to be a transient as was claimed by \citet{vanzella2020b}. As we discuss below, our analysis supports their conclusion with additional evidence.

{The two lens modeling papers mentioned above appeared in the literature after our lensing analysis was finalized. We emphasise that our lens modeling and source analysis, which we present here, were conducted entirely independently from the lensing analysis, source identification, and redshift measurements of \citet{diego22} and \citet{pignataro21}.} Our analysis is based on all the \hst\ imaging available to date; we increase the number of lensed systems that are used as constraints from five to fifteen, and update the redshift measurement of one of these systems.

In this paper, we present a detailed lensing analysis of \clustername\ based on extensive  \hst\ imaging and VLT/\muse\ spectroscopy.
A number of science questions can be addressed with accurate strong gravitational lensing models of clusters of galaxies such as \clustername. 
The strong lensing model maps the projected mass density distribution at the core of the cluster, which can in turn be used to study its structure and the interplay between the dark and luminous components that reside in clusters' deep potential wells. Other model outputs facilitate investigations of the background Universe using the cluster as cosmic telescope. We derive the lensing magnification in this field, with emphasis on the \Sunburstarc\ and the LyC knot. A measurement of the lensing magnification is required for converting the observed to intrinsic properties of lensed sources. To fully understand the morphology of the source, we derive deflection maps with which to construct a qualitative view of the \Sunburstarc's source plane, and constrain the unlensed, intrinsic sizes of unresolved clumps in this galaxy. A prediction of the time delay between images of the source helps interpreting its observed components and could be useful to constraints cosmological parameters if a variable source is found.

The paper is organized as follows. In \autoref{sec:data} we present the  imaging and spectroscopy datasets used in this work. \autoref{sec:arcs} details the lensing evidence. \autoref{sec:model} describes the lens modeling process. We present and discuss our results in \autoref{sec:results}, and summarise our findings in \autoref{sec:summary}.
Throughout this work we assume a flat cosmology with $\Omega_{\Lambda} = 0.7$, $\Omega_{m}
=0.3$, and $H_0 = 70$ km s$^{-1}$ Mpc$^{-1}$. Magnitudes are reported in the AB system unless otherwise stated. 
We adopt a cluster redshift of $z=\zcluster$ and a systemic redshift of $z=\zarc$ for the \Sunburstarc.

\section{Data} \label{sec:data}
\subsection{\hst\ Imaging and Grism Spectroscopy} \label{sec:imaging}
\clustername\ was the target of several \hst\ programs in Cycle 25. GO-15101 (PI: Dahle) obtained multi-band imaging and grism spectroscopy of \clustername\ with the goals of enabling a robust lensing analysis and investigating the physical conditions of the lensed galaxy. That program used five orbits of broad-band imaging in the F555W, F814W, F105W and F140W filters; four orbits of F410M medium band imaging; and five orbits of WFC3 G141 grism spectroscopy. A second program, GO-15377 (PI: Bayliss), complemented these data with four orbits of \hst\ imaging in F606W, F098M, F125W, and F160W  as part of a \chandra\ Cycle-19 program, to determine whether the source galaxy hosts an active galactic nucleus (AGN) and observe the diffuse X-ray gas of the lensing cluster.
Third, the Cycle~25 mid-cycle program GO-15418 (PI: Dahle) used three \hst\ orbits to probe the LyC of the lensed source in F275W. Some of the visits failed due to gyroscope problems, and were repeated.
\autoref{tab:hstobs} tabulates the successful and failed \hst\ Cycle~25 observations from the programs listed above. 

The target was also observed by two programs in \hst\ Cycle~27. A two-orbit integration in F390W was executed in 2020 July 13 by GO-15949 (PI: Gladders) and is used in this work; the remaining data from Cycle~27 will be presented and analyzed in forthcoming publications.

The \hst\ imaging data were reduced following standard procedures, after inspecting each image for quality assurance, given the higher than usual failure rate due to the gyroscope problems in Cycles 25--26. 
We used the Drizzlepac\footnote{\url{http://www.stsci.edu/scientific- community/software/drizzlepac.html}} software package to reduce the data and align the frames to a common reference grid, as follows. First, exposures in each filter that were taken within a single visit were drizzled using the \texttt{astrodrizzle} routine using a Gaussian kernel with a drop size (\texttt{final\_pixfrac}) of $0.8$. Next, for each filter in which observations were executed over multiple visits, the drizzled images from each visit were aligned to a common world coordinate system (WCS) using the \texttt{tweakreg} routine. These WCS solutions were propagated back to the individual exposures 
using \texttt{tweakback} before all exposures in a single filter were drizzled together using \texttt{astrodrizzle}, with the same parameters listed above. Finally, the drizzled images were aligned in WCS space, again using \texttt{tweakreg}, and drizzled using \texttt{astrodrizzle} with the same kernel and drop size onto a common reference grid with North up and a pixel scale of $0\farcs03$ per pixel.

WFC3 IR G141 Grism observations from GO-15101 were executed with two telescope roll angles, \texttt{ORIENT}=$27.37^\circ$ and $355.37^\circ$. The data were reduced using the reduction package \texttt{Grizli}\footnote{ \url{https://github.com/gbrammer/grizli}}, using standard reduction procedures. We used these data to search for or attempt to confirm candidate lensed galaxies in the Grism spectra.

\begin{deluxetable*}{clllll}
\tablecaption{\HST\ observations\label{tab:hstobs}}
\tablewidth{0pt}
\tablehead{
\colhead{Program} & \colhead{Camera} & \colhead{Filter/grating} & \colhead{Date (UT)} & \colhead{Exp. Time [s]}  & 
}
\startdata
GO-15101 & WFC3/UVIS & F410M & 2019-08-13 & 13285 &  \\
 & WFC3/UVIS & F555W & 2019-03-11 & 2852 (*) &  \\
 & WFC3/UVIS & F555W & 2019-03-11 & 2964 (*) &  \\
 & WFC3/UVIS & F555W & 2019-06-24 & 2792 &  \\
 & WFC3/UVIS & F555W & 2019-06-24 & 2824 &  \\
 & ACS  & F814W & 2018-02-21 & 2544 &  \\
 & ACS  & F814W & 2018-02-22 & 2736 &  \\
 & WFC3/IR & F105W & 2019-03-12 & 1312 (*) & \\
 & WFC3/IR & F105W & 2019-06-24 & 1312 &  \\
 & WFC3/IR & F140W & 2019-03-12 & 1312 (*) & \\
 & WFC3/IR & F140W & 2019-06-24 & 1312 &  \\
 & WFC3/GRISM & F140W & 2019-05-12 & 868 &  \\
 & WFC3/GRISM & F140W & 2019-06-30 & 562 &  \\
 & WFC3/GRISM & G141 & 2019-05-12 & 8418 &  \\
 & WFC3/GRISM & G141 & 2019-06-30 & 5612 &  \\
GO-15377 & WFC3/IR & F098M & 2019-03-06 & 1359 &  \\
 & WFC3/IR & F125W & 2019-03-06 & 1359 &  \\
 & WFC3/IR & F125W & 2019-04-30 & 1359 &  \\
 & WFC3/IR & F160W & 2018-09-27 & 1359 &  \\
 & WFC3/UVIS & F606W & 2018-09-27 & 1484 &  \\
 & WFC3/UVIS & F606W & 2019-01-17 & 2982 (*) & \\
 & WFC3/UVIS & F606W & 2019-03-06 & 1484 (*) & \\
 & WFC3/UVIS & F606W & 2019-03-06 & 1484 (*) & \\
 & WFC3/UVIS & F606W & 2019-03-12 & 2922 &  \\
 & WFC3/UVIS & F606W & 2019-04-30 & 1424 &  \\
GO-15418 & WFC3/UVIS & F275W & 2018-04-14 & 6318 &  \\
 & WFC3/UVIS & F275W & 2018-04-08 & 3104 &  \\
GO-15949$^1$ & WFC3/UVIS & F390W & 2020-07-14 & 3922 &  \\
\enddata
\tablecomments{\hst\ observations that were used for the lensing analysis presented in this paper. (*) these visits failed due to gyroscope problems in Cycles 25-26 and subsequently repeated. $^1$Additional data from GO-15949 beyond that listed here were not used in this work.}
\end{deluxetable*}

\subsection{\Chandra\ X-ray Data}
\label{sec:chandra}
The field containing the Sunburst Arc was observed with the {\it Chandra X-ray Observatory} under observation ID 20442. The purpose of this observation was to constrain any bright X-ray emission from the lensed galaxy, while also producing a robust detection of the foreground cluster lens. The observation was executed as a single 39.53 ks exposure with the aimpoint located near the center of the I3 chip in the ACIS-I array. To minimize background the observation was performed in VFAINT telemetry mode. We reduced the {\it Chandra} data using {\it Chandra} Interactive Analysis of Observations (\texttt{CIAO v4.13}) with \texttt{CALDB v4.9.6} to apply routine processing. The data were filtered for flares using the \texttt{lc\_sigma\_clip} function in the \texttt{lightcurves} Python package that is included in \texttt{CIAO}, resulting in a usable integration time of 38.53 ks. We apply a 0.5--7 keV energy filter to the reduced event file and correct for a small ($\sim$1\arcsec) astrometric offset between the {\it Chandra} and \hst\ data by comparing the coordinates of a galaxy that appears both in the \hst\ field of view and as an X-ray point source in the {\it Chandra} data. We then use the 0.5--7 keV {\it Chandra} image to measure the basic X-ray properties of the foreground cluster lens. The X-ray peak (brightest pixel) is at R.A. 15:50:07.053, Decl.  $-$78:11:29.165, and the X-ray centroid at: R.A. 15:50:06.82, Decl.  $-$78:11:29.921, both within $1''$ of the BCG in projection.

The total observed $0.5-7$~keV X-ray flux is $2.54\times10^{-12}$ erg cm$^{-2}$ s$^{-1}$. Excising the central $r\sim150$~kpc to estimate a core-excised flux results in $f_{0.5-7keV }= 1.06\times10^{-12}$  erg cm$^{-2}$ s$^{-1}$.
The ratio of those two values (core-excised to total fluxes) make it clear that the cluster has an extremely strong cool core. For example, a comparison to other Planck clusters analyzed by \citet{Mantz2018} places this cluster on the extreme end in terms of $f_{\rm ce}  / f_{\rm total}$, meaning that its X-ray emission is among the most core-dominated. This is consistent with what is observed in the optical and IR; the BCG appears dust-obscured, with evident filamentary star formation activity both in imaging and spectroscopy, similar to what is seen in other extremely strong cool cores \citep[e.g., the Phoenix Cluster; ][]{McDonald2012,McDonald2019}.

\subsection{Ground-Based Spectroscopy}
\label{sec:spectroscopy}
\clustername\ was observed with  \muse\ on the VLT on 2016 May 13 (ESO program 297.A-2012(A); PI Aghanim). The data were retrieved from the ESO archive; the data reduction process is detailed in \cite{Weilbacher2020} and \cite{Urrutia2019}. The field was observed for 2966 s under good conditions, with $0\farcs7$ seeing, spanning $4750-9300$\,\AA\ with $R\simeq2100$. 
The MUSE cube results in a spectrum for each resolution unit, which is a powerful way to derive redshifts for numerous sources in the field of view, since there is no need to decide in advance which sources to target. We use the \muse\ data to confirm counter images in lensed systems and measure their redshifts.

The field was observed with \magelong\ on the Magellan-I 6.5-m Baade telescope, covering multiple distinct image plane regions in the \north\ and \northwest\ arcs. The details of the observation and data reduction are given in \cite{rivera-thorsen17,rivera-thorsen19,lopez20}, and most comprehensively in \citet[][in prep]{rigby2022inprep}.  These data were used to confirm the identifications of some of the clumps in the sunburst arc.

\begin{figure*}
\epsscale{1.0}
\plotone{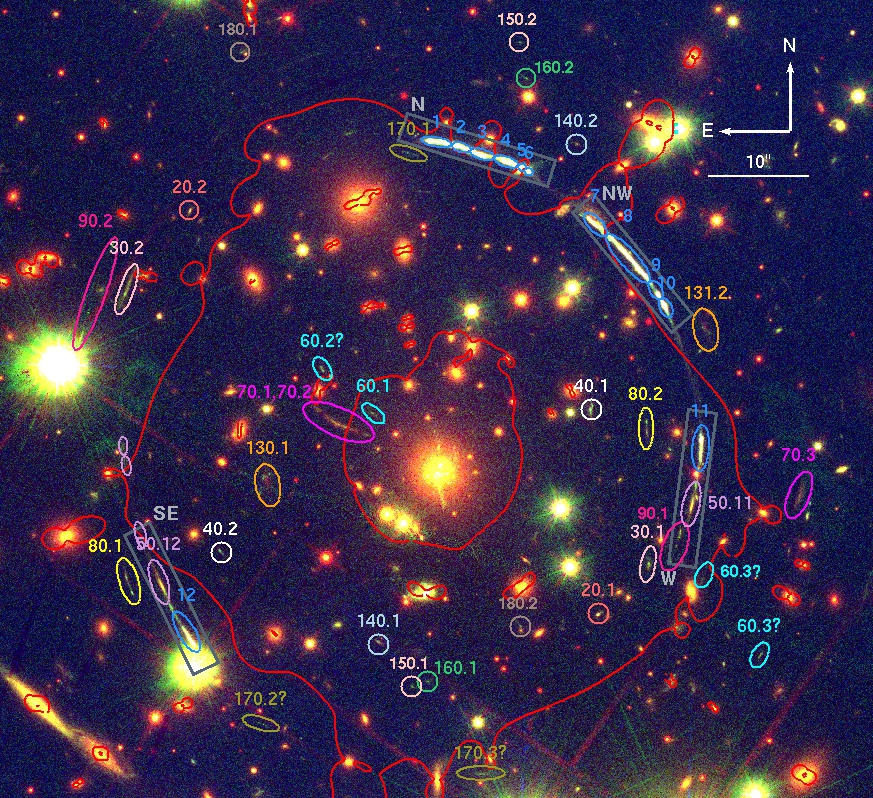}
\caption{Imaging data of \clustername, composed of F140W+F160W (R), F606W (G), F390W (B). The secure and candidate multiple image systems are labeled and color coded; candidate images are annotated with a question mark. The individual clumps within each image are not labeled, to reduce clutter.  The critical curve for a source at $z=\zarc$ is overlaid in red. Gray rectangles mark the giant arcs of the \Sunburstarc; they are shown in more detail in \autoref{fig:arc1}. See also \autoref{fig:stamps} for zoomed-in view of the multiple images.
} \label{fig:fov}
\end{figure*}

\begin{figure*}
\epsscale{1.0}
\plotone{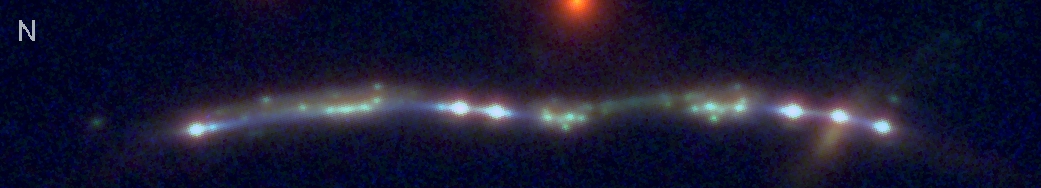}
\plotone{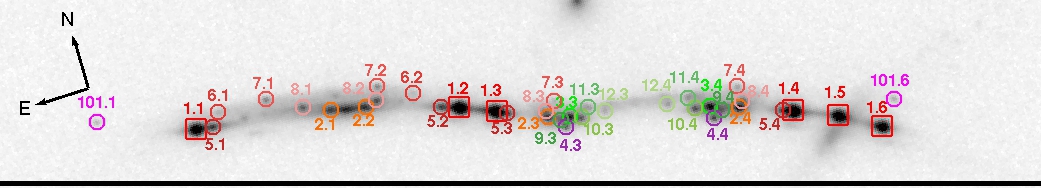}
\plotone{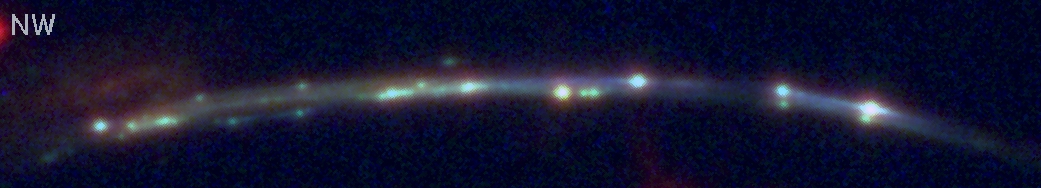}
\plotone{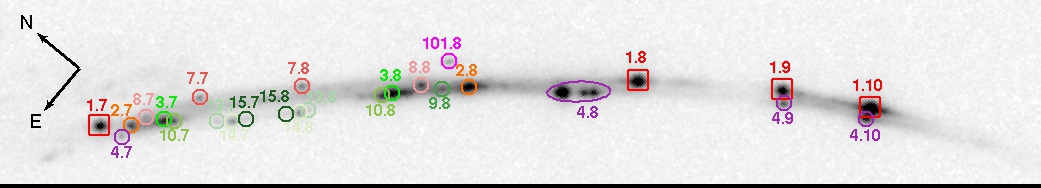}
\plotone{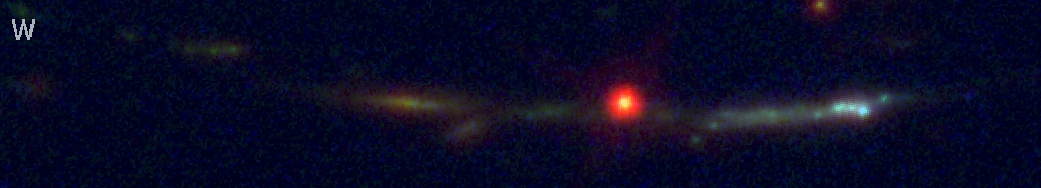}
\plotone{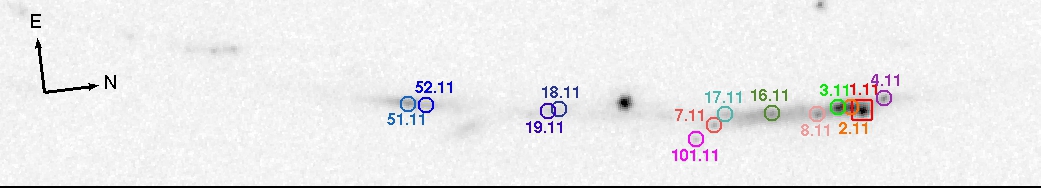}
\plotone{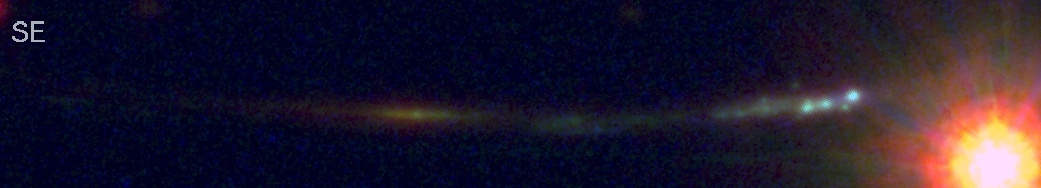}
\plotone{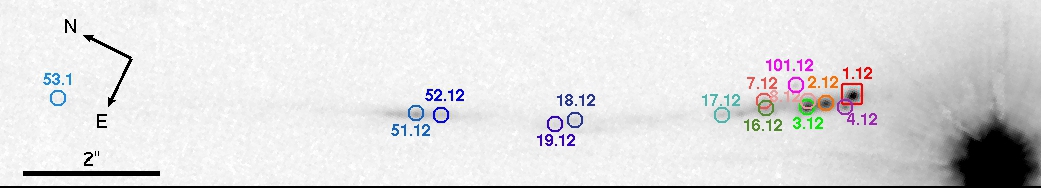}
\caption{Zoomed-in view of the giant arcs of the Sunburst system. The color frames are a composite of F140W (R), F606W (G), and F410M (B), probing the restframe optical, UV continuum, and \lya, respectively. The filters, scaling, brightness, and saturation levels were chosen to emphasize color variations between emission knots.  In each monochrome frame we label the identified emission knots. A $2\farcs0$ scale bar is shown in the bottom frame. The gray boxes in \autoref{fig:fov} indicate the location of the frames shown here within the cluster field of view.
} \label{fig:arc1}
\end{figure*}

\vspace{10pt} 
\section{Lensing Evidence} \label{sec:arcs}
\subsection{The Strong Lensing Interpretation of Source 1: the \Sunburstarc}
The most prominent lensing feature in the field of \clustername\ (\autoref{fig:fov}) is the highly-magnified galaxy discovered by \cite{dahle16}. In the discovery paper, based on ground-based $R$- and $z$-band imaging data from the ESO New Technology Telescope (NTT), \cite{dahle16} reported that the source is lensed into three giant arcs, which appear north, \northwest, and west of the BCG (S1, S2, S3 in their notation), with a possible fourth image to the \southeast.  They measure a spectroscopic redshift of $z= 2.3686 \pm 0.0006$ from nebular emission lines, and $z= 2.3708\pm0.0004$ from the \lya\ emission line. \cite{rivera-thorsen17} determined the systemic redshift of the system as $z=2.37094\pm0.00001$ from nebular line emission in the FIRE spectra.
A systemic redshift of $z=2.37034\pm0.00024$ was obtained from an average of the velocities of the narrow components of the  \OIII\ line in several FIRE pointings (Mainali et al. 2022 in prep). Differences of this order are within expectation due to velocity differences between the galaxy components, e.g., winds and relative motions, and insignificant for the purpose of lens modeling. In this work we adopt $z=\zarc$ as the systemic redshift.

High-resolution \hst\ imaging data are invaluable in understanding the lensing configuration.   
Studying the first F814W and F275W observations, obtained in February and April 2018, respectively, \cite{rivera-thorsen19} found that the \north\ and \northwest\ arcs are each composed of multiple images, and in particular, the bright emission clump in the source is lensed into a dozen images: six in the \north\ arc, four in the \northwest\ arc, and one each in the \west\ and \southeast\ arcs. 
With the multi-band \hst\ imaging available to date, we are able to identify numerous emission knots in each image, and match these emission knots between images. \autoref{tab:mainarctable} and \autoref{fig:arc1} list the positions of the emission knots, and label their mapping between the lensed images.  The identified emission knots within the \Sunburstarc\ are labeled with prefixes 1 through 19. ID 1.xx is assigned to the bright LyC emitting knots identified by \citet{rivera-thorsen19}. The suffix xx denotes the ID of the lensed image within the multiple-image family. For consistency, we follow the multiplicity ID numbers $1-12$ that were introduced by \citet{rivera-thorsen19}. 

In \autoref{fig:arc1} we show the four images of the \Sunburstarc s: \north, \northwest, \west, and \southeast, in color rendition from \hst\ F140W, F606W, and F410M. These bands are selected for this color rendition in order to display the clump-to-clump color variations. For a source at $z=\zarc$, the broad-band filters F140W and F606W sample emission redward of the $4000$\AA\ break, and restframe UV, respectively, while the medium-band filter F410M  is centered on the \lya\ emission. The scale and stretch in this figure are tuned to bring out the clumpiness of the arc and visually resolve its numerous emission knots.
We label and color-code the emission knots in the panel adjacent to each color rendition. Knot 1.x (i.e., the one emitting LyC radiation, hereafter LCE clump) is labeled with a red square in all the arcs. We label the next three brightest knots with 2.x, 3.x, 4.x, and the remaining knots are labeled and color-coded to guide the eye to their projected location and arc-to-arc symmetry rather than by brightness. 

The \north\ arc (arc 1) consists of six partial images of the source (labeled x.1 through x.6 in \autoref{tab:mainarctable} and \autoref{fig:arc1}). In two of the images, x.5 and x.6, only the bright LyC emitting knot is seen. The other four include several more emission knots in addition to the bright knot:  {knots 2, 5, 6, 7, 8 in images x.1 and x.2; and knots 2--12 in images x.3 and x.4}. The unusually high multiplicity within the \north\ arc is due to contributions to the lensing potential from three cluster-member galaxies, which complicate the shape of the critical curve in this region.

The \west\ arc (arc 3) and the \southeast\ arc (arc 4) are both full images of the source. {As such, these are the only instances where the properties of the galaxy as a whole can be measured (Kim et al., in prep)}. Only in these two images do we observe a likely companion galaxy, labeled 50 in \autoref{tab:arcstable} {(see also the next section). The knots associated with it, 51.x, 52.x as well as knots 18.x and 19.x are not observed in the \north\ and \northwest\ arcs due to the lensing geometry. Because these images have a lower magnification, some of the knots that are identified in other arcs are either not resolved or too faint to be uniquely matched in these arcs. We map knots 16.x and 17.x between these two arcs, but note that they could be matched to knots 11.x and/or 12.x in the \north\ arc (see also \autoref{sec:source})}.  

The \northwest\ arc (arc 2) is the least understood from the perspective of lensing geometry. It has at least two, and up to four, partial images of the source. \citet{rivera-thorsen19} compared the MagE spectra of several spatially distinct regions along the arc, including images 1.1, 1.4$+$1.5, 1.8, and 1.10 (images 1, 4$+$5, 8, and 10 in their notation, respectively). They found that the spectral features of these images are indistinguishable, arguing in favor of these being images of the same source,  which implies that the LCE knot is observed with a multiplicity of 4$\times$ in this arc.
While in the \north\ arc the 6$\times$ multiplicity of the LCE knot can be readily explained by the complexity of the lensing potential due to nearby cluster-member galaxies, no such perturbers are observed near the west end of the \northwest\ arc.  

Indeed, near images 1.9, and 1.10, the lensing geometry is not trivial. 
One explanation for the apparent multiplicity in this region, as suggested by \citet{diego22}, is that the critical curve serpentines around image 1.9; this requires a lensing potential that is more complex than can be explained from the observed galaxy distribution, e.g., due to a very low surface brightness or a yet unobserved mass component.
They forced a high multiplicity in this region by constraining the lens model with critical points between images 1.8, 1.9, and 1.10 (h, i, and l in their notation).
The model presented in this paper does not reproduce the multiplicity of images 1.9 and 1.10, because it lacks the flexibility on small scales that is needed in order to wind the critical curve around image 1.9 using only lens components with an observed counterpart. Our early attempts to force the critical curve through those positions (by using critical curve constraints, similar in principle to what was recently done by \citealt{diego22}) resulted in unrealistic magnification ratios between the lensed images. We ruled out those models based on the unrealistic magnifications, and abandoned this direction. We will explore the possibility that a low surface brightness or a ``dark'' clump is responsible for the lensing complexity in this region in future work. Despite its limitation in the southern part of the \northwest\ arc, our lens model correctly recovers the numerous lensing constraints from multiple source planes (\autoref{sec:arcs}) elsewhere; the global lens properties and measurements are not affected.

An alternative explanation of the observed images 1.9 and 1.10 is that the LCE clump itself is composed of multiple smaller clumps, unresolved in the other instances but resolved in this image due to its high tangential distortion. 
Additional data, such as the approved \JWST\ NIRSpec Integral Field Unit (IFU) spectroscopy of this part of the arc (JWST-GO program\,02555, PI: Rivera-Thorsen), will provide more clues.  The IFU spectroscopy of images 1.9 and 1.10 could reaffirm that they are indeed identical to the other images of the LCE, thus confirming its 4$\times$ multiplicity within this arc, and requiring a more complex lens model with sufficient flexibility to wind a critical curve around image 1.9, as was attempted by \citet{diego22}. 
Deep imaging may reveal a faint interloper that complicates the lensing potential, or put limits on the surface brightness of such a component if undetected. The first public \JWST\ data already showcase its power to detect low surface brightness structures (e.g., the intracluster light; \citealt{Mahler22b,Montes2022}).
Alternatively, the IFU data may reveal that one or two of these clumps differ from the main LCE source, indicating that the LCE clump is actually composed of several distinct LyC leaking regions. 

A further anomaly observed in the \northwest\ arc is a bright, unresolved, emission clump that displays significantly different colors and spectral features, compared to other clumps of similar brightness. This emission clump can be seen in \autoref{fig:arc1}, as the bright clump in the left part of the 4.8 ellipse. 
Because a counter image with similar brightness does not appear in any of the other giant arcs, in particular those which display full images of the source, \cite{vanzella2020b} argued that this unresolved emission clump \citep[labeled \texttt{Tr} in][]{vanzella20a} is a transient. 
They further identified Bowen fluorescence lines \citep{Bowen1934} in its spectrum. \citet{diego22} examined the  possibility that the observed brightness and morphology of this candidate (which they nicknamed ``{\it Godzilla}'' in their paper) is due to an unobserved lensing perturber in close proximity. They found that such a perturber requires mass of order $\sim10^8$\msun. 
In Section \ref{sec:weirdclump}, we discuss the discrepant emission clump in more detail, and demonstrate that it is unlikely to be a transient. 

\begin{figure*}
\epsscale{1.0}
\plotone{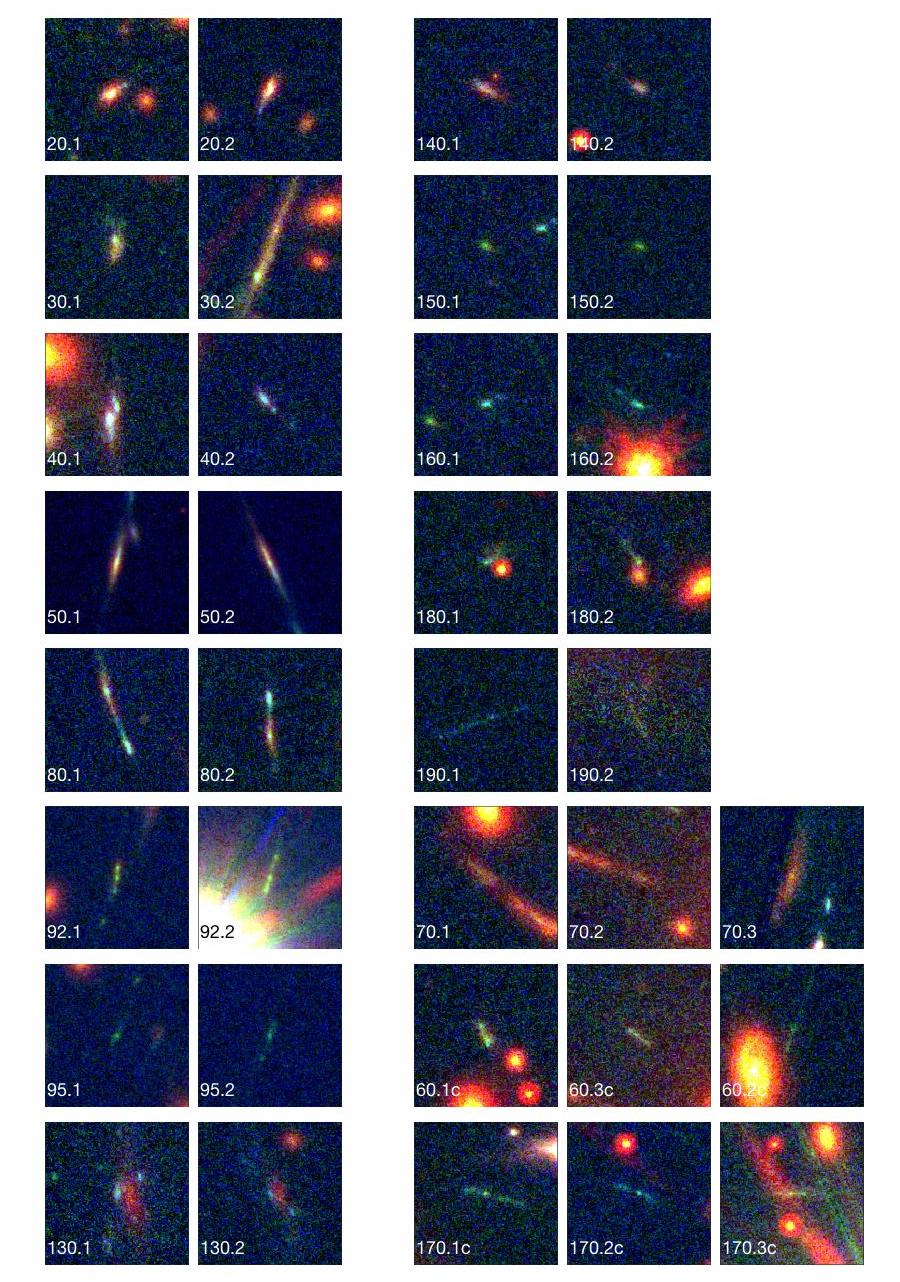}
\caption{Close-up view of the multiple images of strongly lensed galaxies identified in the field of \clustername. The color image is composed of F140W+F160W (R), F606W (G), F390W (B), chosen to highlight color gradients within the lensed galaxies to show their internal structure, and color differences between galaxies.  North is up, and each image is $4\farcs0\times 4\farcs0$.
} \label{fig:stamps}
\end{figure*}

\vspace{10pt} 
\subsection{Secondary lensed systems}
From the multi-band \hst\ imaging and archival MUSE spectroscopy, we  identify multiple images of several other galaxies that are also strongly lensed by \clustername. Their positions, and where available, spectroscopic redshifts, are used as constraints in the lensing analysis. To avoid confusion with the IDs of clumps within the \Sunburstarc, we assign these secondary lensed galaxies ID numbers starting at 20. The tens digit denotes the source ID, and the ones digit denotes the label of each emission knot. The digit right of the decimal point labels each image within the multiple image ``family'' of the same source. For example, images 21.1 and 21.2 are two lensed images of clump number 1 in source number 20. Candidate features have an additional prefix $c$. \autoref{fig:fov} shows the locations of the images of the secondary systems. The positions of individual emission knots, and the available redshifts, are tabulated in \autoref{tab:arcstable}. In \autoref{fig:stamps} we present a close-up view of the identified multiple images of each system. We note that for each of the tangential systems the model predicts a demagnified ($0<|\mu|<1$) counter image behind the BCG. These highly demagnified images are predicted to be several magnitudes fainter than the observed arcs, and as such they are not expected to be detectable in the current data. 
As noted in \autoref{sec:intro} and \autoref{sec:chandra}, the BCG area is active with star formation, which is expected for a cool core cluster such as \clustername. Since star formation chains near the BCG can mimic the appearance of radial arcs \citep[e.g.,][]{Sharon2014,Tremblay2014,McDonald2019}, we treat the central region with extra care in order to not misidentify star formation as lensed features.

\paragraph{Source 101} Likely associated with the \Sunburstarc, Source 101 (\autoref{fig:arc1}) appears as a single emission knot projected less than half an arcsecond from the main arc, but with only five multiple images: two by the \north\ arc, one by the \northwest\ arc, one by the \west\ arc and one by the \southeast\ arc. A sixth image is predicted by the northern tip of the \northwest\ arc, and we tentatively identify it blended in the light of a foreground galaxy. The sixth image candidate is not used as a constraint. 
Due to its proximity to the main arc, and the lack of independent redshift measurement, we cannot spectroscopically rule out its association with the main arc. When leaving its redshift as a free parameter, the model-predicted redshift for this source converges to that of the main arc. Furthermore, narrow-band \HST\ imaging reveals strong line emission from all the images of Source 101 at a wavelength consistent with [O~III]~4959\,\AA\ at the redshift of the \Sunburstarc\ (Rigby et al., in prep.). We therefore proceed by assigning this clump the same redshift as the \Sunburstarc. 

\paragraph{Source 20} This source appears with two images, one southwest and one northeast of the BCG. Although we do not have a spectroscopic redshift of this galaxy, the morphology and color variations along the images provide confidence in their identification as multiple images of the same source. 
\paragraph{Source 30} We detect two images of source~30, with similar morphology and color. A single emission line appears in both locations in the \muse\ data, placing this source at $z_{spec}=2.460$, assuming this line is from C~III]. The secure lensing identification, and high-confidence redshifts of other sources rule out other possible emission lines, which would result in too high or too low redshifts (e.g., \lya, [OII] [OIII]). The east image of this source (30.2), as well as image 2 of source~90, appear in a region with higher tangential shear than the west images of the same sources (\autoref{fig:fov}). 
\paragraph{Source 40} Source 40 is lensed into two images, projected 17 and 22 arcseconds from the cluster center. Archival MUSE spectroscopy indicates that it is a star-forming galaxy at $\zspec=1.1484$, based on several emission lines that are observed in both of its images, 40.1 and 40.2. 
\paragraph{Source 50} We identify two images of a galaxy, one extending the \west\ \Sunburstarc\ to the south, and the other extending the \southeast\ \Sunburstarc\ to the east.  Images of this source do not appear near any of the other \Sunburstarc\ images because of the location of the source with respect to the caustics in the source plane; this is consistent with expectation from the lensing geometry. Source~50 is smoother and redder than the main arc, with a distinctive color gradient. We identify two clumps within the images to be used as constraints. 

\muse\ spectroscopy of images 50.11 and 50.12 places this source at the same redshift as the \Sunburstarc, $\zspec=2.3709$, based on weak C~III] emission lines and Si~II and C~IV in absorption in both images of the source. 
We note that the spectrum of 50.11 contains features from an interloping galaxy at $\zspec=0.7373$, based on a bright double emission feature at $6477.0$\AA\ and $6482.1$\AA\ from [O~II] $3737$ doublet, corroborated by other lines including H$\beta$ emission and [O~III] $5008$ emission. 
We identify the interloper as a faint galaxy in close projected proximity, $0\farcs94$, \northwest\ of 51.11. The identified interloper has the same spectroscopic features that contaminate the \muse\ spectrum of 50.11. 

Our redshift analysis of source 50 disagrees with the redshift reported for this galaxy by \citet{pignataro21}, who measured $z=2.393$ (Sys-3 in their paper). While their paper does not provide information regarding the spectroscopic features that were used to establish this redshift, it could possibly be based on a misidentification of the [O~II] line from the $z=0.7373$ interloper, as a C~III] line at  $z=2.393$. 
Source~50 is close in projection to the  \Sunburstarc\ ($\sim 6$~kpc away in projected distance in the source plane according to our lensing analysis, \autoref{sec:source}), which could explain the absorption lines had source 50 been in the background. 
Nevertheless, the identification of the  C~III] emission line adds confidence to our $\zspec=2.3709$ interpretation.

A close inspection of the imaging data reveals a thin elongated arc that extends 50.12 towards the northeast. This emission is likely the magnified image of just the edge of Source~50, which is lensed to this location by contribution to the lensing potential from a nearby cluster-member galaxy. The three images that make this extension are labeled 53.1, 53.2, 53.3.

\paragraph{Source 60 (candidate)} We identify this image as a candidate radial arc. At this time, we have no spectroscopic redshift for it. There are several possible counter images at the expected locations that match its surface brightness, colors, and expected lensing shear, but neither can be spectroscopically confirmed as counter images. We therefore do not include this system as a constraint in the lens model.
\paragraph{Source 70} This source appears as a radial arc, observed east of the cluster core. We identify a candidate counter image of this arc, west of the \west\ \Sunburstarc. The candidate has similar colors to the radial arc. The direction in which the arcs are sheared is consistent with expectation from lensing geometry. The radial arc appears in the \MUSE\ data with low significance, and a redshift could not be secured from these data. The location of the counter image falls outside of the footprint of the archival \MUSE\ data, and we are therefore unable to confirm or rule it out with the \MUSE\ data. 
Similarly, the arc is too faint in the Cycle 25 GRISM data available to this study. It is possible that the \HST-WFC3/G280 data (GO-15966) are deep enough for spectroscopic redshift confirmation. The redshift of this source is left as a free parameter in our modeling. 
\paragraph{Source 80} The two images of source 80 present distinctive color and morphology, which secure their identification despite not having a spectroscopic redshift confirmation. 
\paragraph{Source 90} Source 90 has two images, one appears south of the \west\ \Sunburstarc, and the second one appears \northeast\ of the cluster core. We measure the redshift of both images, $\zspec =3.5053$, based on \muse\ detection of extended \Lya\ emission coincident with this galaxy. We identify five unique emission knots in the \hst\ imaging of this source, while the \lya\ emission appears much more extended than the optical/IR emission. 
The eastern image of source~90 spans $10\farcs6$, about three times more extended than its west counterpart, at $3\farcs4$, indicating that there is significantly more lensing shear in the east region. We observe similar behavior in the nearby system~30. 
\paragraph{Source 130} System~130 has two images, both resolved, with a red center and two distinct blue emission knots. We use the blue emission knots as constraints, as their location can be more precisely determined than that of the extended red center. We have no spectroscopic redshift for this system, but its unique morphology and colors provide confidence in its identification.
\paragraph{Sources 140, 150, 160} 
Each of these systems has two images, with no spectroscopic redshift. We identify them based on their colors and morphology.
\paragraph{Source 170 (candidate)} We identify three images of this candidate system, with similar surface brightness and morphology, which are predicted by the lens model to be counter images of each other. Image 170.1 appears in the north part of the field, close to image 1.1. The lens model predicts its counter images in the south of the field. We identify arcs with similar morphology, that match the lensing parity and shear direction that are predicted by the model. We note that image 170.1 appears slightly greener than 170.2 in the rendition shown in \autoref{fig:stamps}, likely due to contamination from bright nearby cluster galaxies. The colors of 170.3 are contaminated by a nearby star. To be conservative, these candidate images were not used as constraints.
\paragraph{Source 180}
We identify two images of this source, one in the northeast and one southwest of the BCG.  The \muse\ spectra of both images suggest a low confidence line, which could be CIII] at $z=2.582$. Due to its low confidence the suggestive redshift is not used as constraint. The morphology of the two images is similar, and consistent with the expected lensing parity and shear direction.
\paragraph{Source 190}
We identify two faint candidate images of this source, a radial arc \southwest\ of the BCG, and a counter image \northwest. 

\startlongtable
\begin{deluxetable*}{lllrll} 
\tablecolumns{4} 
\tablecaption{List of lensing constraints within the \Sunburstarc\label{tab:mainarctable}} 
\tablehead{\colhead{ID} &
            \colhead{R.A. [deg]}    & 
            \colhead{Decl. [deg]}    & 
            \colhead{$\mu$ } &          
            \colhead{1$\sigma$ uncertainty on $\mu$}       
            \\[-8pt]
            \colhead{} &
            \colhead{J2000}     & 
            \colhead{J2000}    & 
            \colhead{}       & 
            \colhead{[lower upper]}             }
\startdata 
\input{planckarc_arc1_mag}
\enddata 
\tablecomments{The IDs and positions of emission knots in the \Sunburstarc,  identified in the multiple images of the source and used as lensing constraints. 
   Clumps are labeled as $A.X$ where $A$ is a number of the clump, and $X$ is a number or a letter indicating the ID of the lensed image within 
the multiple image family.  The best-fit model-predicted magnifications for a point source located at the exact position of each feature are given in the fourth column, and the brackets indicate the lower and upper magnification corresponding to 1$\sigma$ confidence interval in the parameter space, sampled from the MCMC chain. (*) The magnifications of images 1.9, 1.10, 4.9, 4.10 are measured at their observed positions; they should be used with caution, as the lens model does not correctly recover these positions. The redshift of the \Sunburstarc, $z=\zarc$, was measured by 
 \cite{dahle16}, \cite{rivera-thorsen19}, and \cite{rivera-thorsen17}, and confirmed with \muse\ spectroscopy by this work. $^1$The redshift of source 50 was measured by this work, placing it at the same redshift as the main arc.  $^2$Source 100 is possibly associated with the same source as the \Sunburstarc, and is either part of the same galaxy or a companion (see \autoref{sec:results})}
\end{deluxetable*}

\startlongtable
\begin{deluxetable*}{llllll} 
\tablecolumns{7} 
\tablecaption{List of lensing constraints from secondary lensed galaxies \label{tab:arcstable}} 
\tablehead{\colhead{ID} &
            \colhead{R.A. [deg]}    & 
            \colhead{Decl. [deg]}    & 
            \colhead{$z_{spec}$}     & 
            \colhead{$z_{model}$}       & 
            \colhead{Notes}       \\[-8pt]
            \colhead{} &
            \colhead{J2000}     & 
            \colhead{J2000}    & 
            \colhead{}       & 
            \colhead{}       & 
            \colhead{}             }
\startdata 
\input{planckarc_arcs}
\enddata 
\tablecomments{The IDs, positions, and redshifts of lensed multiply-imaged galaxies other than the \Sunburstarc, that were used as constraints in this work. Where possible, individual emission knots in each image are identified and used as lensing constraints. 
The IDs of images of lensed galaxies are labeled as $AB.X$ or $AAB.X$ where $A$ or $AA$ is a number indicating the source ID (or system name);  $B$ is a number indicating the ID of the emission knot within the system; and $X$ is a number indicating the ID of the lensed image within the multiple image family. A prefix $c$ identifies candidates. The reference for spectroscopic redshifts is this work; see \autoref{sec:arcs}. The redshifts of source 40 and 90 are consistent with the independent measurement of \citet{pignataro21} using the same data, of 1.186, 3.505, respectively. 
}
\end{deluxetable*}

\section{Strong Lens Modeling Procedure} \label{sec:model}
\subsection{Methodology}
We use the lens modeling software \lenstool\ \citep{jullo07}, which is a parametric lens modeling algorithm, i.e., it assumes that the lens plane can be constructed from a linear combination of individual parametric mass halos. In this work, we model each mass component with a pseudo-isothermal ellipsiodal mass distribution \citep[PIEMD, also known as dPIE;][]{eliasdottir07}, with seven parameters: position ($x$, $y$, measured in arcseconds from a reference point), ellipticity $e$, position angle $\theta$, core radius $r_c$, truncation radius $r_{cut}$, and a normalization $\sigma$. 

Following procedures that have become standard in the field, the lens plane is modeled iteratively. The most obvious lensing evidence (i.e., the most prominent features in the \Sunburstarc) is used to constrain an initial lens model. The model is then used to identify additional lensing constraints, and predict the locations of their counter images. We use the archival \MUSE\ data to spectroscopically confirm candidates. Following the recommendations of \cite{johnson16}, we re-initiate the modeling process when including new spectroscopically confirmed constraints, in order to not bias the lens model by its own predictions. As new constraints are identified, we  increase the lens model complexity as needed by either freeing the parameters of galaxies in close proximity to the giant arc, or by adding dark matter halos that represent correlated substructure \citep[e.g.,][]{mahler18}. 

This lens model is a significant improvement upon the model published by \citet{rivera-thorsen19}, which only used the \Sunburstarc\ images as constraints, and consequently had a limited flexibility with which to model the lens plane. Two important datasets enabled this improvement: multi-band \hst\ imaging to identify lensed sources and resolve substructure within lensed images, and \MUSE\ spectroscopy to derive reliable redshifts.
The large number of constraints coming from the clumpy, highly magnified \Sunburstarc\ and the other lensed sources discussed in \autoref{sec:arcs} enables construction of a lens model with high flexibility and complexity, allowing a large number of free parameters. The availability of constraints from secondary systems, i.e., multiply-imaged galaxies at different source planes, provides important leverage for constraining the global properties of the cluster mass distribution, and lowers the uncertainties \citep{johnson16}.

The components of the lens model are described in the next section. The model uses positional constraints from 146 images of 48 identified clumps within 15 multiply imaged lensed galaxies (19 of the clumps are in the \Sunburstarc) with multiplicities ranging from 2--12, and has a total of 57 free parameters including 9 free redshifts. The best-fit model RMS scatter between predicted and observed images is $0\farcs85$.

\subsection{Lens Components}
The lens plane is represented by one cluster-scale and two group-scale dark matter halos, supplemented by galaxy-scale halos. We position the cluster-scale halo near the BCG. We free all the parameters of this halo with the exception of the cut radius, which is fixed at $1500$ kpc, as for cluster scale halos this radius is far beyond the observed lensing evidence. This mass component accounts for the dark matter halo of the cluster, as well as the hot X-ray gas, which we find to be centered within $1''$ of the BCG.
We place one group-size halo in the general direction of a small group of bright cluster galaxies approximately $25$ arcseconds \northeast\ of the BCG, and one group-size halo $41$ arcseconds south of the BCG fixed to the position of a luminous cluster member. All the other parameters, of both halos are left free, with broad priors.
The cluster halos are supplemented with \Ngal\ galaxy-scale halos, positioned on the observed locations of cluster-member galaxies. The selection procedure of these galaxies is explained in the next section. 
The galaxies are modeled as PIEMDs, with positional parameters fixed at their observed properties (R.A., Decl., ellipticity, and position angle), and the other parameters scaled to their luminosity following a parametric scaling relation given in \citet[][eqn. 28]{limousin05}, with pivot parameter $M(L^*)=19.45$ mag.
Several galaxies were modeled as individual halos, allowing \lenstool\ to solve for their best-fit parameters. These are either galaxies that lie in close projected proximity to the arcs, or galaxies that are not expected to follow the same scaling relations as the red-sequence cluster galaxies, e.g., the BCG, and spectroscopically-confirmed star forming galaxies at the cluster redshift. 

At the core of the cluster, the UV/optical light of the central galaxy appears to be obscured by dust. We therefore measure its position in the reddest band, F160W, which should be the least affected by dust obscuration. We leave all the parameters of the BCG free, including its position, with a $\pm0\farcs5$ prior around the F160W centroid. 

We include in the lens model a lensing galaxy which is positioned close to the northern part of the \northwest\ arc, behind the cluster, at $z_{spec}=0.733$ \citep[galaxy G1 in][]{lopez20}. We approximate this interloper as contributing to the lensing potential at the same plane by allowing high flexibility in modeling its mass. The degeneracy between the normalization and the distance term means that the mass of this galaxy cannot be computed reliably with this approximation. Similarly, a faint interloper galaxy near image 50.11 ($\zspec=0.737$; see Section~\ref{sec:arcs}) is also included in the lens model at the cluster redshift. We refer the reader to \citet{Raney2020} for a thorough examination of the  implications of the modeling approach of projecting the interloper onto the same lens plane of the main lensing structure.

\subsection{Selection of Cluster Galaxies}
The mass associated with cluster member galaxies contributes to the lensing potential of the cluster. 
We use Source Extractor \citep{sextractor} in dual-image mode to generate a photometric catalog of the field, with the F814W image used as the detection image, and \texttt{MAG\_AUTO} measured in both the F814W and F606W images within the same aperture. We use a detection threshold of 5 sigma, and deblending contrast of 0.001. Stars were flagged by their locus in \texttt{MU\_MAX} vs \texttt{MAG\_AUTO} plane and excluded from the catalog. 
We select cluster-member galaxies based on their F814W-F606W color in a color-magnitude diagram, following \cite{gladdersyee2000}. These particular bands were selected because they span the characteristic $4000$\AA\ break of passive galaxies at the cluster redshift. The ACS data provide the widest field of view around the cluster core.
Finally, we visually inspect the galaxy catalog and remove objects that were erroneously picked as cluster members, such as diffraction spikes, faint stars, and parts of over-deblended galaxies.

\section{Results and Discussion} \label{sec:results}
\subsection{Cluster Mass Distribution}\label{sec:mass}
\begin{figure}
\epsscale{1.2}
\plotone{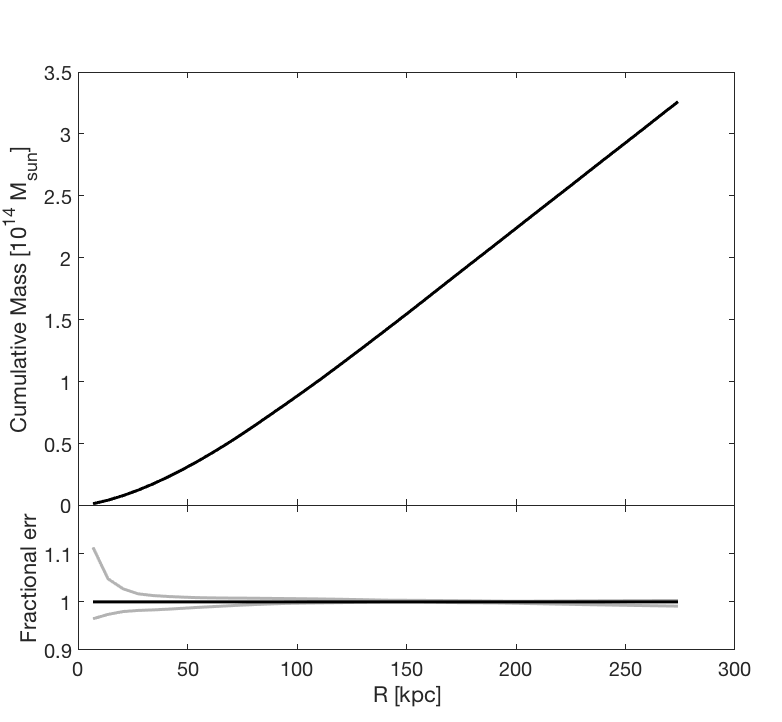}
\caption{The cumulative projected mass density profile at the core of the lensing cluster \clustername. The functional $1-\sigma$ uncertainty (statistical) is shown in the bottom panel.} \label{fig:mass}
\end{figure}

\autoref{fig:mass} shows the projected mass density profile of \clustername, measured as a function of distance from the BCG. 
We find that the projected mass density of the core of \clustername, enclosed within $R=250$ kpc is  $M(<250 {\rm kpc})= $ \clustermass. 
For comparison with previous mass estimates, we report the mass within $40'' = 228.3$ kpc, $M(<40'')= 2.629 ^{+0.005}_{-0.015} \times 10^{14}$\msun, and the mass within 200 kpc, $M(<200 {\rm kpc})=2.238 ^{+0.003}_{-0.007}\times 10^{14}$\msun, which is similar to what was estimated by \citet{pignataro21} and \citet{diego22}. It is unclear whether there is a discrepancy since no uncertainties were provided in those publications.
Interestingly, we find that the mass within the radius of the giant arc, $M(<169 {\rm kpc}) = 1.809 ^{+0.005}_{-0.002}\times 10^{14}$\msun, is in perfect agreement with what was estimated by \citet{dahle16}, $M_E = 1.8\pm0.6\times 10^{14}$\msun\ for $r_E=169\pm25$ kpc. Their mass estimate is based on the relationship between the Einstein radius and the projected mass enclosed within it, $M(<r_E)=\pi r^2_E\Sigma_{crit}$, for a circularly symmetric lens. Such a close agreement between the detailed lens model measurement and the mass enclosed by the Einstein radius method indicates that the circular symmetry assumption is valid for this cluster \citep{remolina21}.

The main cluster halo coincides with the BCG and the X-ray gas centroid to within $1''$, which is consistent with \clustername\ being a relaxed cluster.

\subsection{Lensing Magnification}\label{sec:magnification}
\begin{figure*}
\epsscale{1.0}
\plotone{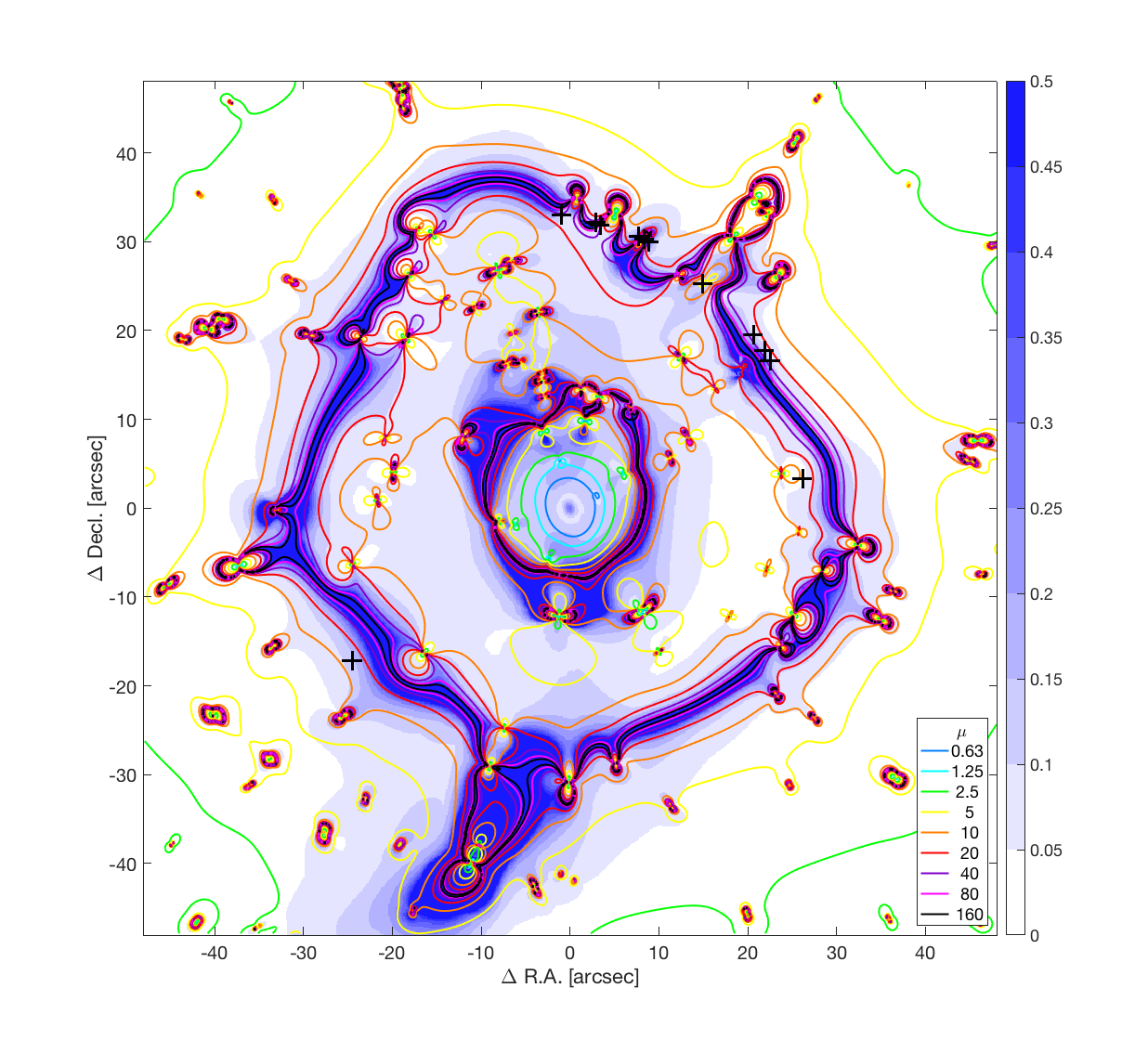}
\caption{The magnification for a source redshift $z=\zarc$ is shown in contours. The fractional uncertainty is represented by the shaded colormap. The uncertainty $\sigma$ is estimated from steps in the MCMC chain, approximately indicating where 68\% of the results fall within $\mu\pm\sigma\mu$. The positions of the 12 images of the LyC knot are marked with `+' symbols. The image coordinates are measured from R.A.=237.5294767, Decl.=$-$78.19167258. See \autoref{fig:magnificationzoom} for a zoom on the \north\ arc.} \label{fig:magnification}
\end{figure*}

\begin{figure*}
\epsscale{1.0}
\plotone{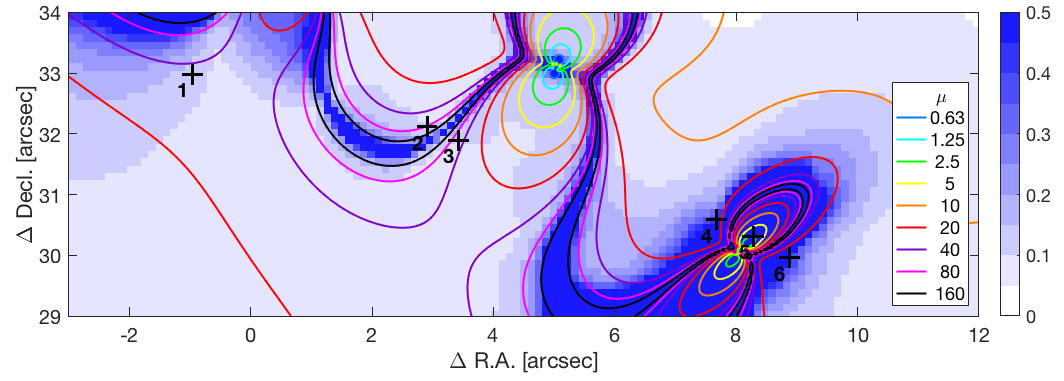}
\caption{Same as \autoref{fig:magnification}, zoomed in on the \north\ arc (images 1--6 of the source). North is up, East is left.} \label{fig:magnificationzoom}
\end{figure*}

\autoref{fig:magnification} maps the absolute value of the lensing magnification and its statistical uncertainty for a source at $z=\zarc$, with a zoom-in on the \north\ arc shown in \autoref{fig:magnificationzoom}. Consistent with the visual interpretation of the lensing evidence, we find that the \north\ and \northwest\ images of the \Sunburstarc\ form in regions of highest magnification; the \west\ and \southeast\ arcs have the lowest magnification. 

\begin{figure}
\epsscale{1.2}
\plotone{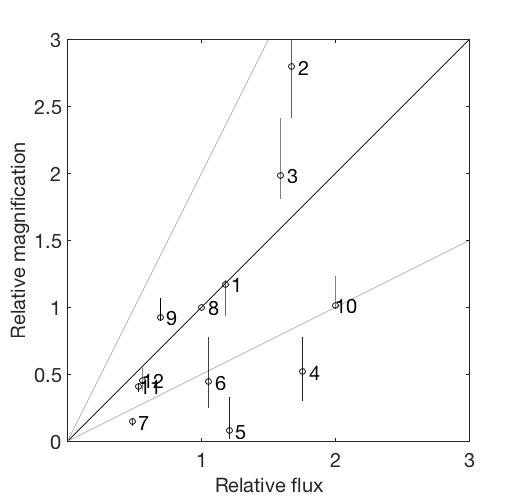}
\caption{Model predicted magnification of the LCE images divided by the predicted magnification of image 1.8, compared to the relative fluxes, measured within a $0\farcs12$ aperture in the F555W image. The solid black line indicates the 1:1 ratio, and gray lines show a factor of 2 deviation. The magnifications of images 1.9, 1.10, are measured at their observed positions; they should be used with caution, as the lens model does not correctly recover these positions.} \label{fig:relative_magnification}
\end{figure}
Estimated over the entire arc, the average magnifications of the \west\ and \southeast\ arcs are $<\!\!\mu_{\rm W}\!\!>=13.5 ^{+2.4}_{-1.0}$ and
$<\!\!\mu_{\rm SE}\!\!>=13.1 ^{+1.0}_{-0.4}$, respectively. Since these are the only two complete images of the source galaxy, a magnification measurement of the entire galaxy is only possible for these two arcs. 
The average magnifications of the  \west\ and \southeast\  arcs are computed by defining an aperture in the image plane, ray-tracing the aperture to the source plane, and dividing the image-plane area by the source-plane area of the ray-traced aperture. For small enough regions that are far from the critical curves, the average magnification is not different from the value given by the magnification map at the center of the feature of interest. 
\autoref{tab:mainarctable} lists the magnification for each of the knots in the \Sunburstarc, measured at the center of each feature. 
To estimate the statistical modeling uncertainties, we generate $\sim100$ lens models, each with a set of parameters from the MCMC chain, which sample a $1-\sigma$ confidence interval in the parameter space. We then run our calculation on the $\sim100$ magnification and deflection outputs.

As a test of the lens model, we compare the model-predicted magnifications to the observed flux of the LCE clump images. The absolute magnification prediction can be tested only when the absolute luminosity of the source can be estimated \citep[e.g., for a standard candle,][]{rodney15}. Since the intrinsic fluxes of the sources in this field are unknown, we evaluate predicted relative magnifications against relative fluxes. We choose to run this analysis on the images of the LCE clump, which is the brightest lensed feature with the highest multiplicity in the field. Fluxes are measured within a $0\farcs12$ aperture in the F555W image, after matching it to the point spread function (PSF) of the F160W filter (a full description of the aperture photometry measurement will be presented in Kim, J. K. et al. 2022, in preparation).  We divide the flux of each image of the LCE by that of image 1.8. Since the magnification uncertainties of the different images are likely correlated, we calculate the magnification relative to that of image 1.8 in each of the $\sim 100$ models sampled from the MCMC, and combine these measurements to estimate the uncertainty on the relative magnification. The formal uncertainty of the flux measurement is negligible compared to the lensing uncertainty, even in cases where the light from the source is somewhat obscured by foreground galaxies. The comparison is shown in \autoref{fig:relative_magnification}. Most of the images have a relative magnification within a factor of 2 from the relative fluxes, within errors. The relative magnifications of images 4, 5, 6, and 7, are lower than expected. This could be due to their proximity to foreground galaxies whose local contribution to the lensing potential is not adequately reproduced by the modeling process. The results of this exercise emphasize that there are systematic uncertainties that are unaccounted for in the lensing analysis, as was repeatedly highlighted in the literature \citep[e.g.,][]{rodney15,zitrin15,johnson16,priwe17,meneghetti17,mahler18,remolina18,kelly18}. Based on this evaluation, we conclude that our statistical magnification uncertainties likely underestimate the full uncertainty by a factor of 4-5. It also indicates that the strong lens model can be improved upon by tapping into relative magnification constraining power, which can provide leverage over the second derivative of the lensing potential. For example, flux anomalies in lensed quasars were found to be indicative of substructure in the lens plane \citep{bradac2002}.  

\subsection{Time Domain}\label{sec:time}
The arrival time surface, also known in the literature as the Fermat potential, is described by the following equation \citep[e.g.,][]{Schneider85}: 
\begin{equation}\label{eq.dt}
\tau(\vec\theta,\vec\beta) = \frac{1+z_l}{c}\frac{D_{l}D_{s}}{D_{ls}}\bigg[\frac{1}{2}(\vec\theta-\vec\beta)^2-\psi(\vec\theta)\bigg],
\end{equation}
where $\vec\beta$ is the position of the source in the source plane, $\vec\theta$ is a coordinate
in the image plane, $z_l$ is the lens redshift, $D_{l}$ and $D_{s}$
are the angular diameter distances from the observer to the lens and to the source,
respectively, $D_{ls}$ is the angular diameter distance from the lens to the source,
and $\psi$ is the lensing potential. Multiple images of a strongly-lensed source form in stationary points in this surface, i.e., minima, maxima, and saddle points. The time delay between any two images can be calculated as the difference in arrival time, $\Delta\tau = \tau(\vec\theta_1,\vec\beta)-\tau(\vec\theta_2,\vec\beta)$. The source location $\vec\beta$ is formally the same for all images of the same source; however, since lens models have finite accuracy, the calculated source positions of the different images are expected to have a small scatter in the source plane. It is therefore common to use the average or magnification-weighted average of the $N$ model-derived source positions of the $N$ multiple images as $\vec\beta$.

\begin{figure}
\epsscale{1.3}
\plotone{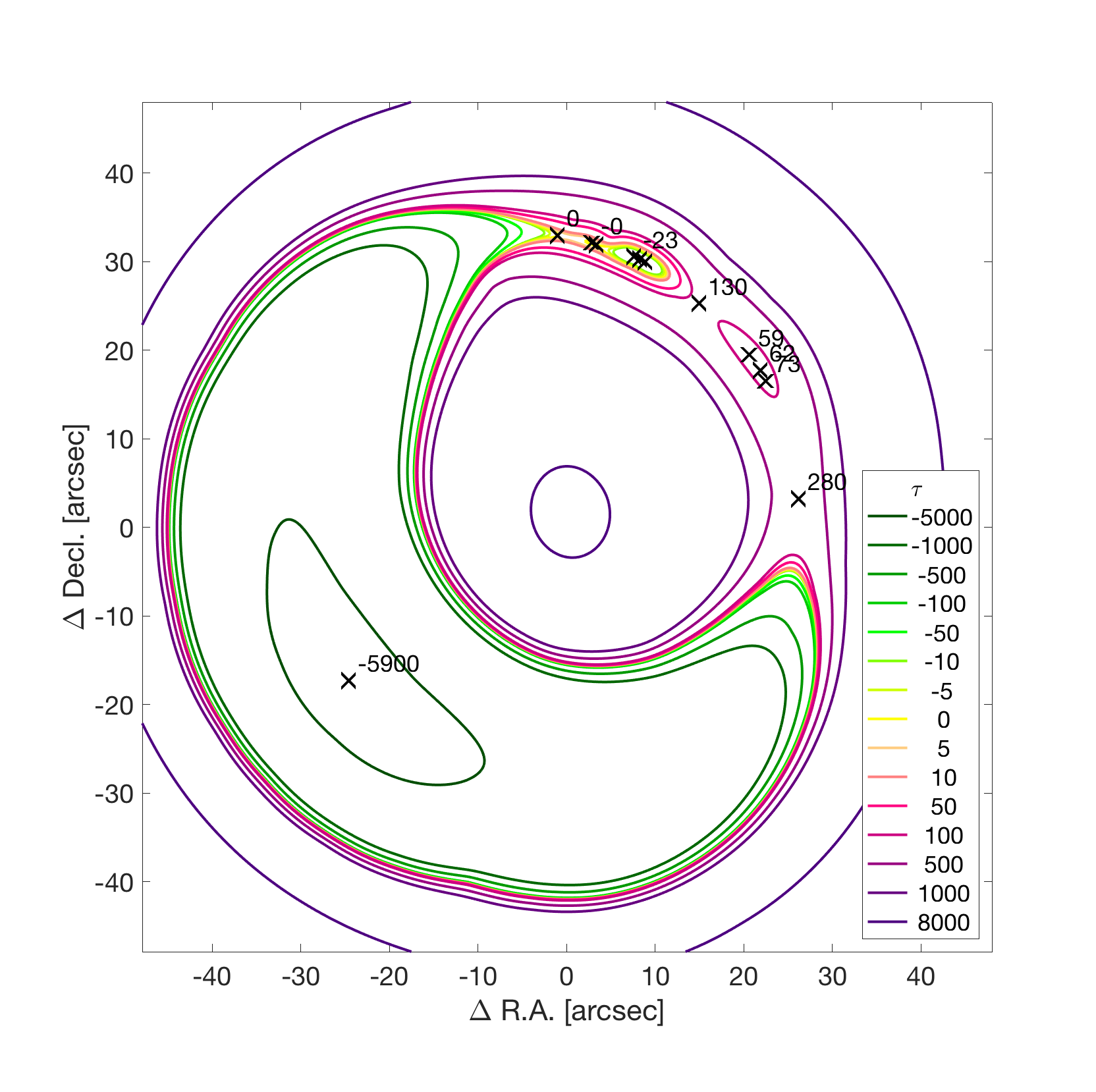}
\caption{The arrival time delay, $\tau$ in observed days (Fermat Potential) relative to image~1 of the LCE knot. Images of the same source position form at stationary points in the Fermat potential (minima, maxima, and saddle points). The \southeast\ arc image forms first (Image 12), followed by the \north\ arc (Images 1--6), the \northwest\ arc (Images 7--10) and \west\ arc (Image 11).} \label{fig:fermat}
\end{figure}

Figure~\ref{fig:fermat} shows the Fermat potential result from the best-fit lens model. The relative arrival time is calculated with respect to image 1.1 of the LCE. We find that a packet of light emerging from the galaxy in the source plane arrives at the \southeast\ arc first (image 12), preceding the other arc positions by 16--17 years. Next to appear are the images of the \north\ arc (1-6), followed by the \northwest\ arc (7-10) and the \west\ arc. Finally, the light from the source plane is predicted to arrive at the location of the (unobserved) demagnified central image some 26 years after the \north\ arc, owing to gravitational time delay by the deep potential well of the cluster.
Interestingly, we find that images 1--11 of the source occur within the span of several months to a year, a result that is qualitatively quite robust to the details of the lens model. This relatively short time delay implies that counter images of transient events in the \Sunburstarc, such as supernova explosions, could potentially be observed, and their time delays measured, within the expected lifetime of current observational facilities. We discuss this point further in the next section.

\subsubsection{The discrepant clump: variability or something else?}\label{sec:weirdclump}
As noted in \autoref{sec:arcs}, a bright point source appears in the \northwest\ arc, with similar brightness but different spectral energy distribution from the LCE images. A counter image of this point source is not apparent in any of the other arcs. \citet{vanzella2020b} interpreted the occurrence of this source to be due to time variability, and postulate that this source is transient in nature. They estimate its magnification at $20\leq \mu \leq 100$, based on lensing symmetry arguments \citep{vanzella20a}.  
Multi-band \hst\ imaging of \clustername\ span more than two years, from 2018-02-21 to 2020-12-30, and while not designed as a cadenced survey, the field was imaged with a wide-throughput filter between 2--4 times each year, with the longest gap of approximately one year between 2019 June 30 and 2020 July 14. During this time, the discrepant clump does not show significant variability. A quantitative epoch-to-epoch comparison is challenging, since different filters were used in different epochs with the exception of F140W that was repeated four times (2019 May, and 2020 Aug, Nov, Dec). Nevertheless, the brightness of the discrepant clump remains qualitatively similar to that of the LCE for over two years, and does not fade away. Moreover, \citet{vanzella2020b} note that the distinct spectral features of this source are observed in \muse\ data as early as 2016. If indeed this discrepant emission is transient in nature, its lightcurve appears to be flat on a several-year observed time scale, and more than a year intrinsic time scale (cosmological expansion stretches the rest-frame time by a factor of $1+z=3.37$), significantly longer than most supernovae.  

\begin{figure*}
\epsscale{1.17}
\plotone{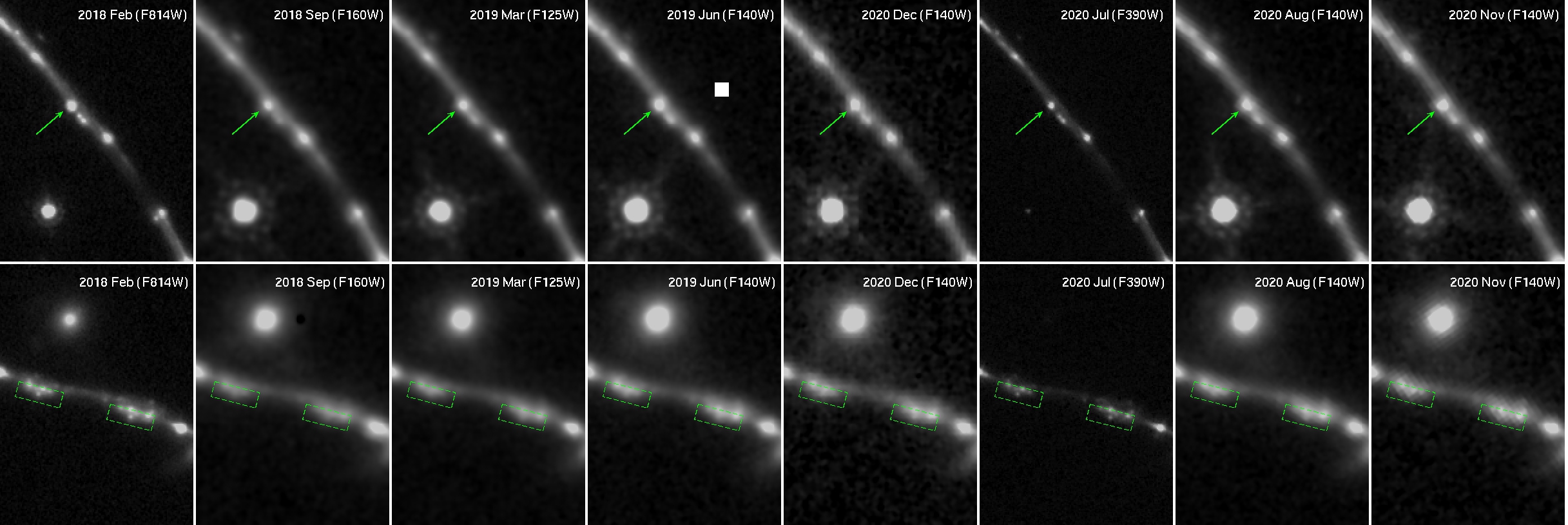}
\caption{Time series imaging to test for variability. The top row shows the region around image 8 of the source and the discrepant point source (green arrow) that was interpreted as transient by \citet{vanzella2020b}. The bottom row shows the region in images 3 and 4 in the \north\ arc where counter images of the discrepant source are predicted to appear within a window of a few months prior to image 8; the discrepant source would be expected to be comparable in brightness to the LCE clump.   
The brightness of the discrepant source is stable throughout the observing window of two years, and we do not observe variability in the \north\ arc, in particular, there are no new or fading sources as bright as the LCE. Cutout frames are $4\farcs5\times6\farcs1$, and North is up.} \label{fig:variability}
\end{figure*}

Our lensing analysis points to a short time delay between the \north\ and \northwest\ arcs, meaning that a transient event in any of these arcs should appear in the other arcs within a few months; time variability alone cannot explain why multiple images of the discrepant source are not detected. Unless this image experiences extreme magnification at the position of the discrepant clump, its counter images should be detectable in close proximity to images 4.3 and 4.4 in the \north\ arc. In \autoref{fig:variability} we show the location of the discrepant arc (top row) and the location of where a counter image should appear (bottom row) in the reddest available \hst\ broad-band filter in each of eight epochs. The discrepant source is clearly detected in each epoch in the \northeast\ arc, with observed brightness comparable to that of the LCE clump. A dashed box marks a broad region around the expected location of the counter image, near clump 4. {If the transient hypothesis were correct, a counter image of the discrepant clump should be visible in these regions with brightness similar to that of the LCE.}  However, {throughout the observing window}, no such sources appear in or disappear from these regions. The lensing analysis indicates that image 7 and 11 of the source, the northern tip of the \northwest\ arc, and the \west\ arcs, respectively, lag less than a year behind image 8 and the discrepant clump. Although these images are less magnified than image 8, a clump with comparable brightness to the LCE clump would have been detectable. We observe no counter image of the discrepant clump in either location.

To summarize, the combined model-predicted time delays and the observational data tell us that the fact that the discrepant clump appears only in the \northwest\ arc must be lensing related and not variability of the clump. Had it been only due to variability, we should have observed its counter image.

The only explanation that remains for why the discrepant clump appears in the \northwest\ arc but is absent from all other counter images of the source galaxy is an extreme magnification at this image-plane location. Such extreme magnification can occur if the source of emission is intrinsically small, and located extremely close to the source-plane caustic. For a point-like source, source-plane caustics formally map to loci of infinite magnification. Such a lensing geometry was invoked in order to explain candidate extremely-magnified single stars due to caustic crossing \citep{kelly18} or proximity \citep{welch22}. 
The average tangential magnification of image 8 is higher than the other arcs, evident by its overall higher distortion; for example, compare the image-plane distance between the three brightest clumps, 1, 2, and 3, in each of the images. In image 8, a grazing critical curve could be resolving clump 4 into two emission knots plus the discrepant clump, the latter being further magnified by its proximity to the critical curve. 

An interesting solution was recently proposed by \citet{diego22}; they explore the possibility that the critical curve near image 8 is perturbed by a small mass, $\sim 10^8$\msun, with no observed counterpart. They show that such a perturber could place the source of the discrepant clump right on top of a caustic, thus magnifying it extremely, and cause image 4.8 to split into the two observed clumps. A similar approach was also employed by \citet{mahler22}, where supermassive black holes were shown to be a plausible reason for lensing anomalies. Upcoming \JWST\ imaging might be able to detect a faint perturber in this location. 

\vspace{10pt} 
\subsection{Sunburst Galaxy Source Plane Analysis} \label{sec:source}

\begin{figure*}
\epsscale{1.15}
\plotone{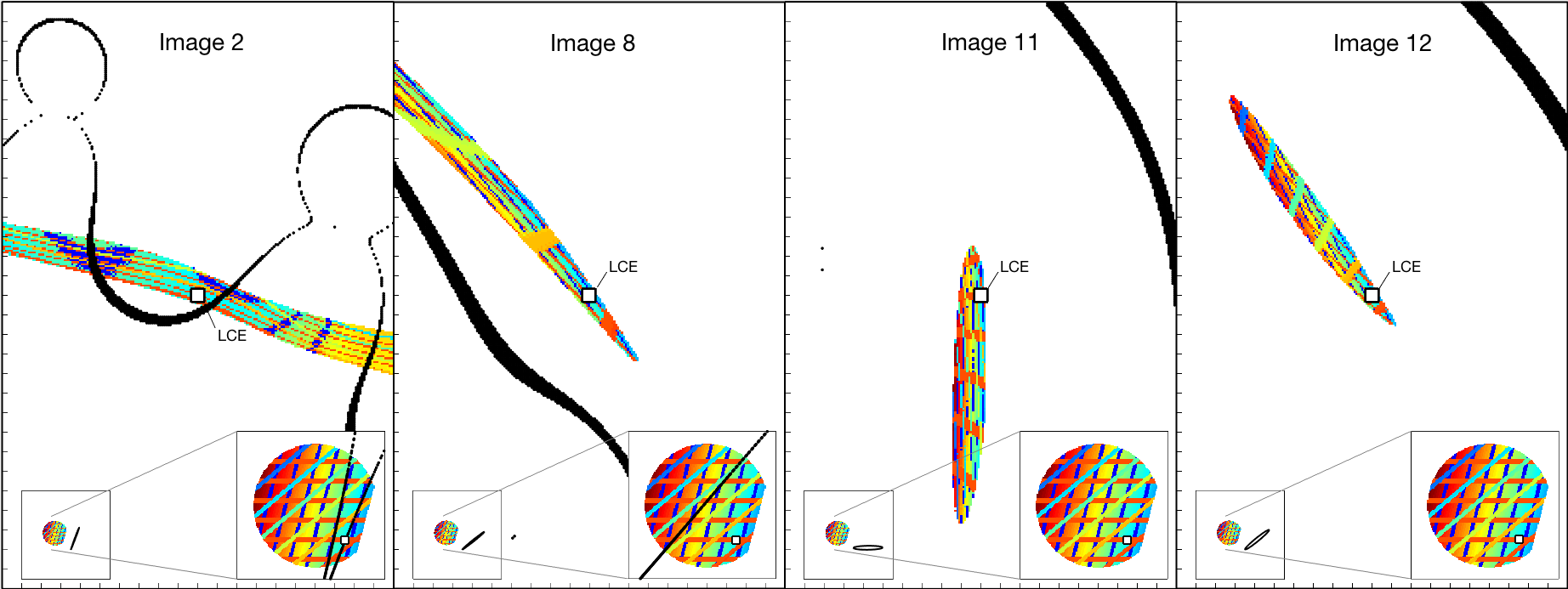}
\caption{An illustration of the different directions in which the source is resolved. The bottom-right inset in each panel shows a zoomed-in view of a mock source, generated to show three main angles. Red lines mark the north-south direction ($90^{\circ}$ ccw), blue lines and background color gradient are plotted at $-15^{\circ}$ and the color sequence lines are at $130^{\circ}$. The left inset shows the mock source at its actual source size. The source is ray-traced through the best-fit model to produce its images, shown in each main panel. Also over-plotted as white box is location of clump 1 (i.e., the Ly-C emitter; LCE), and in black the critical curves (in the main panels) and the relevant source-plane caustics (in the zoom-in panels). The black ellipse illustrates the so-called lensing PSF, which depends on the image location. It is generated by lensing a circle from the image plane to the source plane. The direction of the semi-minor axis of the ellipse is the direction of highest magnification. One can see that at the location of Image 2 (as well as the adjacent images, 1 and 3), the source is stretched in the direction that resolves the blue lines; in Image 8 and 12, the highest resolution is in the color sequence line direction; whereas in image 11 the red lines are resolved. Images 2 and 8 are more highly distorted, as can also be seen from the shape of the lensing PSF ellipse. Finally, by comparing the background color gradient of the source and image of each panel, one can see that only a portion of the source reaches the image plane in images 2 and 8.  
} \label{fig:confusogram}
\end{figure*}

\begin{figure*}
\epsscale{0.45}
\plotone{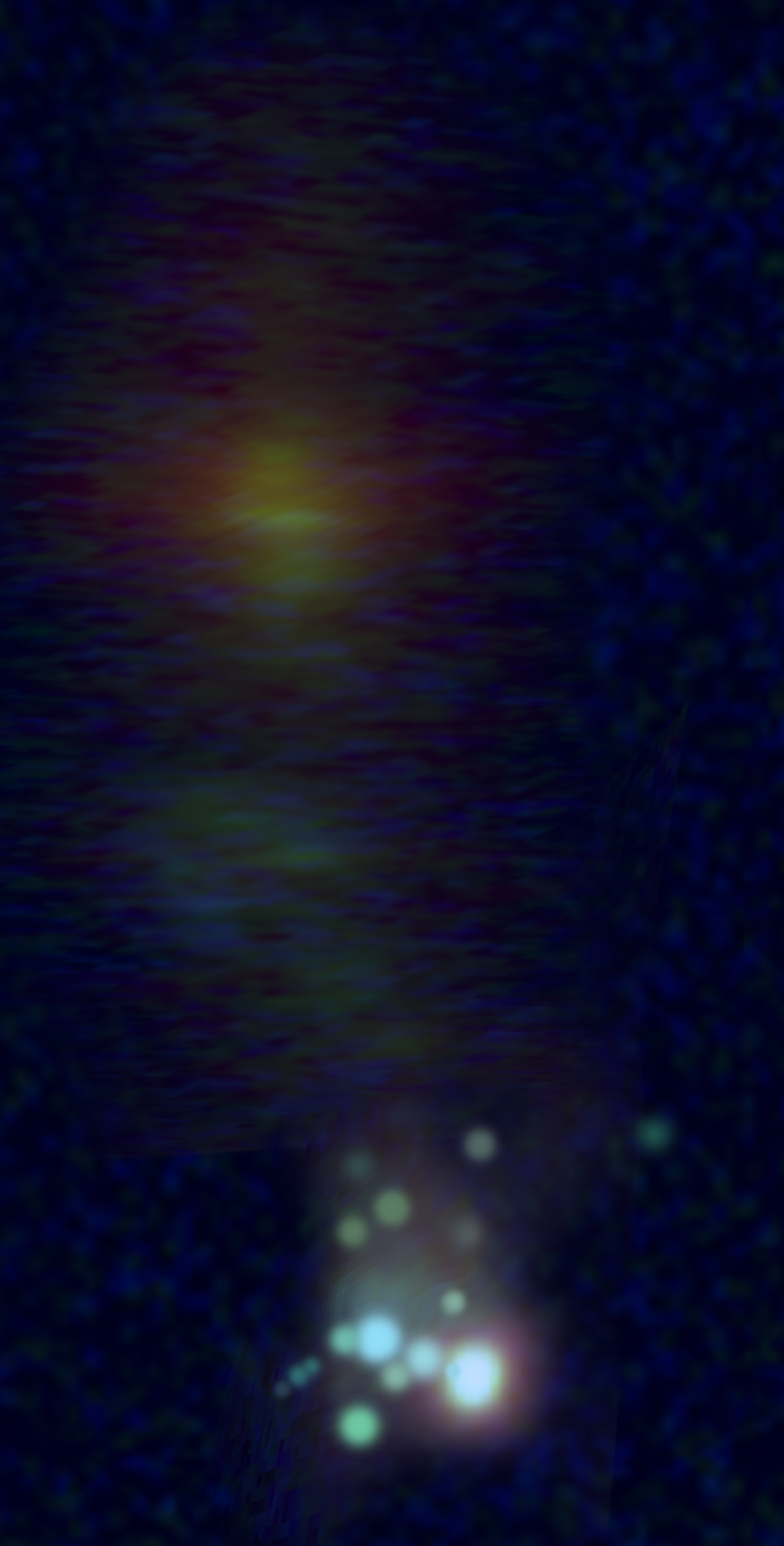}
\plotone{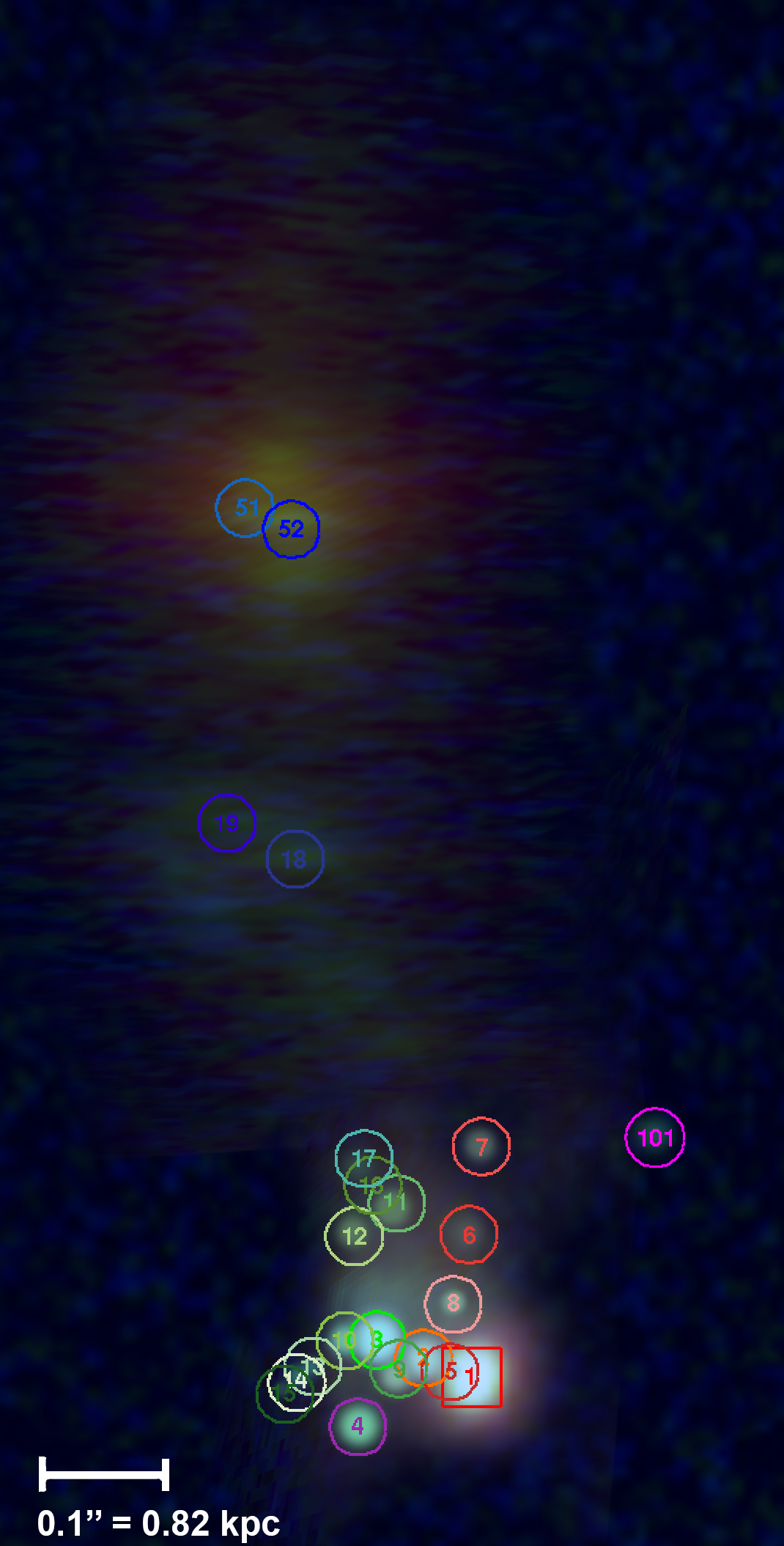}
\caption{An artist's impression approximate view of the source galaxy of the \Sunburstarc. The figure was created by positioning circular light sources in source plane locations as determined by ray-tracing the image plane through the lens model. The color of each light source matches the composite color view in \autoref{fig:arc1}. The emission shown north of the clumpy source is depicted by tracing the rendered image of the \southeast\ arc to its model-predicted source plane position. This includes the image of Source~50 and a ``bridge'' between the two galaxies, containing clumps 18 and 19. 
In the right panel, we label each light source to indicate which image-plane emission clump it is lensed to.
This figure is shown as an approximate visualization, and is not a rigorous flux-calibrated source reconstruction.
} \label{fig:source}
\end{figure*}

\begin{figure*}
\epsscale{0.5}
\plotone{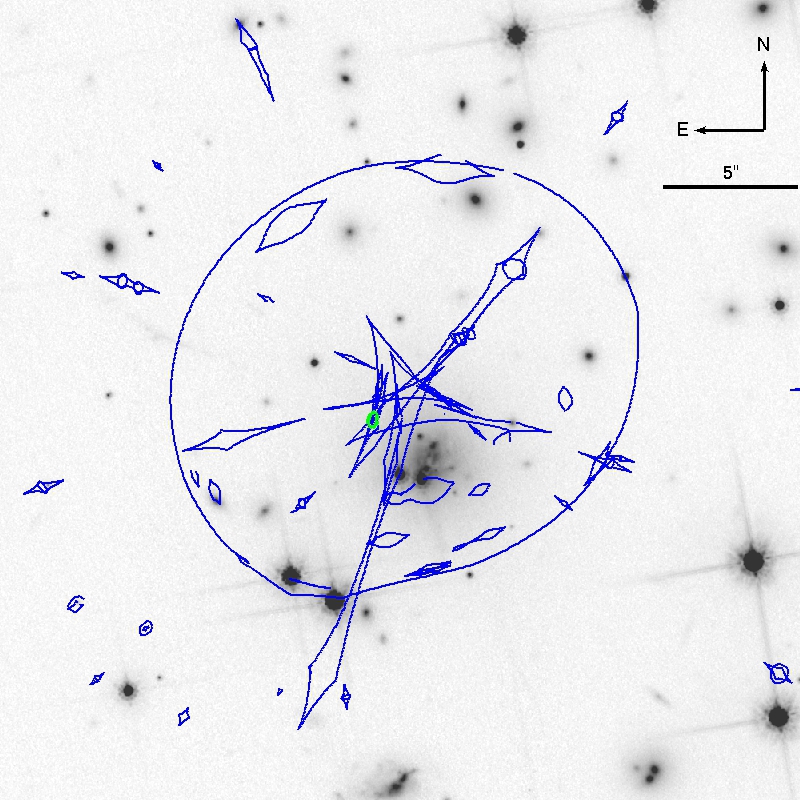}
\plotone{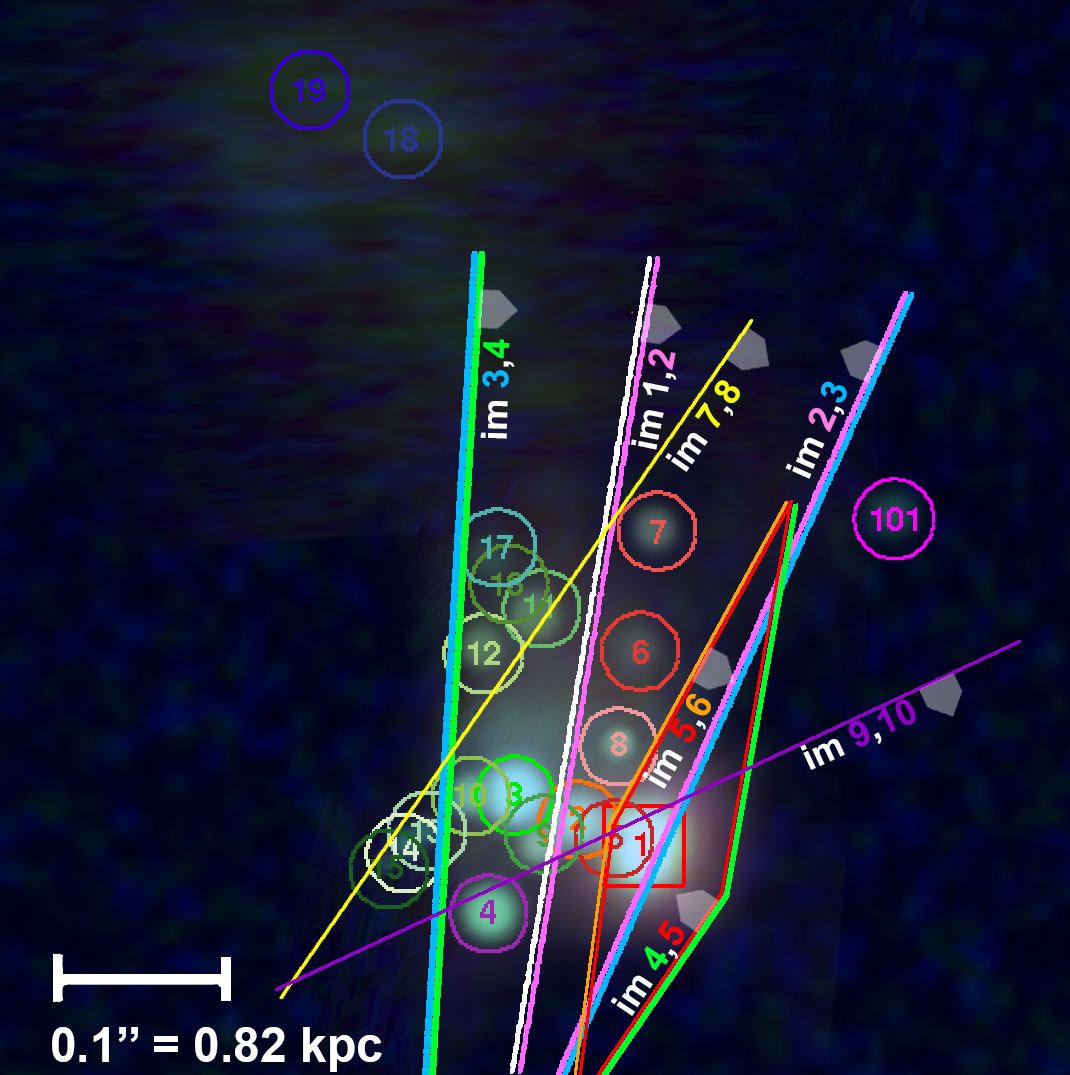}
\caption{\textit{Left:} the lensing potential of \clustername\ forms a complex set of caustics in the source plane. Shown here are the caustics, which are the projection to the source plane of regions of highest magnification in the image plane, for a source at $z=\zarc$. The location of the \Sunburstarc\ source galaxy is shown in green oval. \textit{Right:} zoom-in on the source galaxy. The background is the same as the right panel of \autoref{fig:source}, the labeled artist's impression of the clumpy source galaxy. Onto it, we overlay schematically our best understanding of where the source plane caustics bisect the galaxy, and their positions relative to the identified clumps. In the image plane, each of the 12 images of the \Sunburstarc\ is bound by one or two critical lines, which map to the shown caustic lines in the source plane; clumps that are outside of the boundary caustics of a particular image would not appear in that image location. This figure is shown as an approximate visualization. The typical curvature of the caustics is not depicted in this figure. A more complex critical curve that winds around images 9 and 10 of sources 1 and 4 would produce another caustic, south of the purple line.} \label{fig:caustics}
\end{figure*}

\begin{figure*}
\epsscale{1.2}
\plotone{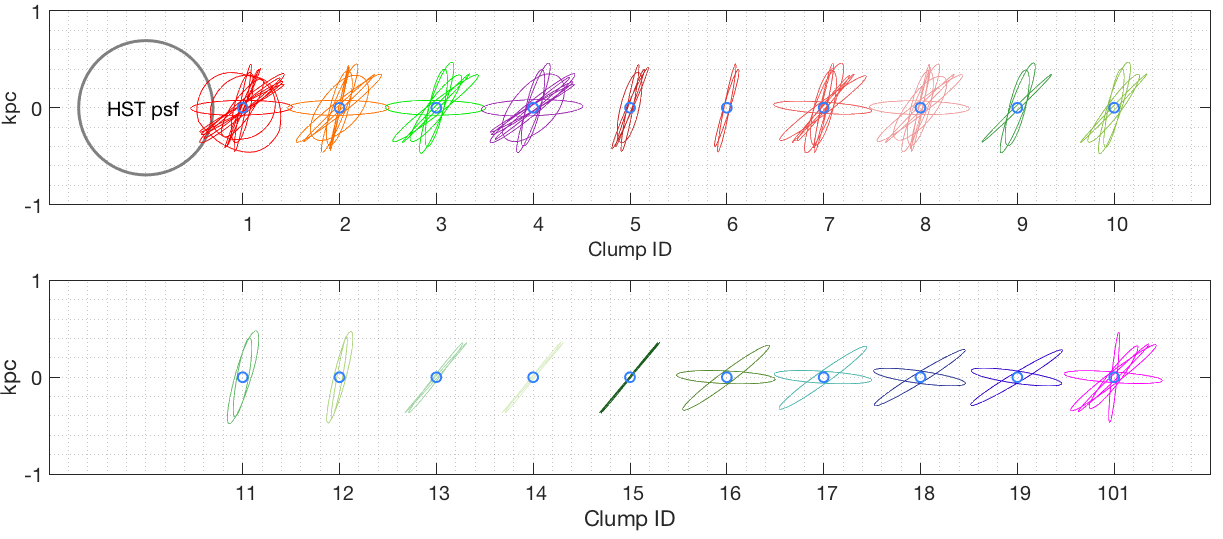}
\caption{Demonstration of the tomographic power of the \clustername\ lens, and an upper limit on clump sizes from the \hst/ACS PSF. We show the source plane ray-traced \hst/ACS F814W PSF at the image-plane location of each of the images of the 19 clumps we identify in the \Sunburstarc. The colors are the same as in \autoref{fig:arc1}. The background grid lines are separated by $0.2$~kpc. The blue circle has a radius of $0.05$~kpc. The unlensed \hst\ PSF, $0\farcs08=0.69$~kpc is shown on the left for comparison.} \label{fig:psf}
\end{figure*}

\begin{figure}
\epsscale{1.3}
\plotone{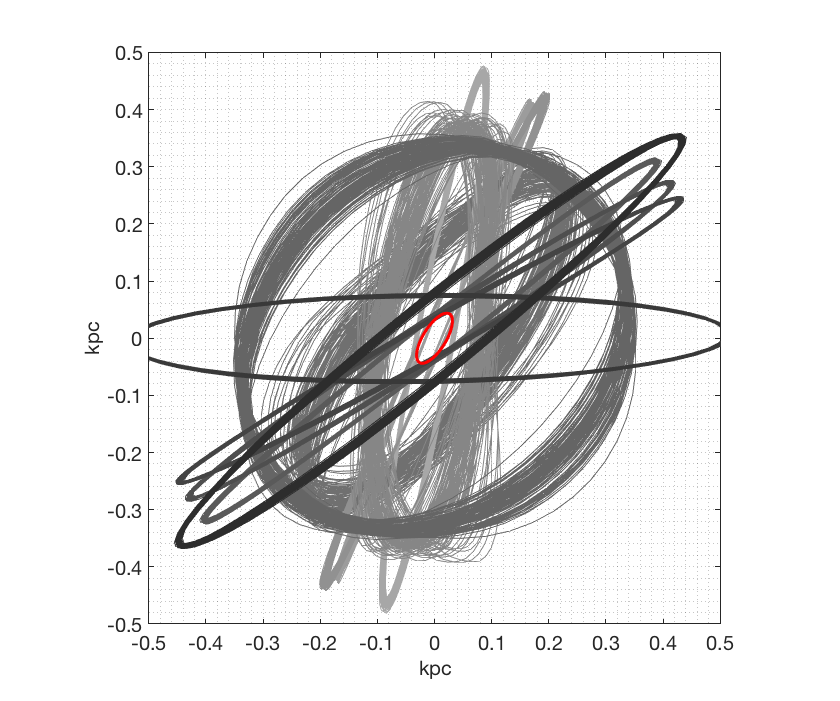}
\caption{An upper limit on the source plane size of the LCE clump, from the \hst/ACS PSF and the lensing magnification. The \hst/ACS F814W PSF is ray traced from the image-plane location of each of the images of the LCE to the source plane. The background grid lines are separated by 0.02~kpc. The gray lines show the projections from 100 models which sample the parameter space, indicating statistical uncertainty; the projections of the high uncertainty image 1.5 is not shown for clarity. The multi-directional magnification constrains the physical size of the LCE to 
$r_{\rm eff}\lesssim31.6$ pc, 
represented with a red ellipse at the center.} \label{fig:psf_LCE}
\end{figure}

The overall effect of gravitational lensing magnification is in amplifying the solid angle of a source while conserving its surface brightness, thus increasing the amount of light that reaches the observer from the source by a factor of $\mu$. The angular amplification is not spatially uniform; it is a combination of an isotropic magnification (convergence, $\kappa$) and distortion (shear, $\gamma$). $\kappa$ and $\gamma$ are combinations of second partial derivative components of the lensing potential, and thus are a function of the image plane position \citep{narayan96}.  

The extreme tangential shear that makes the images of the \Sunburstarc\ acts to resolve  the source galaxy in multiple directions; in each arc, the shear acts most strongly in a preferential primary orientation. We illustrate this behavior in \autoref{fig:confusogram}, showing a mock source and its resulting lensed images in four regions of the image plane: Image 2 in the \north\ arc, Image 8 in the \northwest\ arc, Image 11: the \west\ arc, and Image 12: the \southeast\ arc. While the source does appear to be  magnified in the radial direction, most of its magnification is in the tangential direction. The mock source is designed to highlight the three primary source-plane angles in which the shear acts in this line of sight, with parallel lines plotted at  $-15^\circ$, $90^\circ$, and $130^\circ$ (measured ccw from North), perpendicular to the primary shear directions, $75^\circ$, $0^\circ$, and $40^\circ$, respectively. 
By lensing the mock source through the lens model, we find that the shear near the  \west\ arc (image 11) acts to resolve the source in the north-south direction, separating the red lines in the mock source, which are angled at $90^\circ$. Near the \southeast\ arc (image 12), the shear resolves the source in the northeast-southwest direction, increasing the separation between the color-sequence lines that are plotted at a $130^\circ$ angle. It acts in a similar direction near Image 8. Finally, near Image 2 in the \north\ arc, the resolution is almost perpendicular to that of Image 11, acting to separate the blue parallel lines that are plotted at $-15^\circ$. 

The multi-directional gravitational lensing distortion can therefore enable a tomographic reconstruction of the source plane, facilitating a high resolution analysis, by carefully combining information from differently distorted images. Structures that may appear unresolved in one image of the source may be resolved in a different image. Understanding the directional distortion also enabled an informed selection of our upcoming \JWST\ NIRSpec IFU pointings, to resolve the same source component in different directions (JWST-GO program\, 02555, PI: Rivera-Thorsen). 

A comprehensive computed tomography (a.k.a. CT) reconstruction and forward-modeling of the source plane of the \Sunburstarc\ is reserved for future work. \textit{In this section}, we use the lens model and our understanding of the multi-directional lensing distortion of the source to qualitatively construct the galaxy in the source plane by hand. 
\autoref{fig:source} shows an ``artist's impression'' of the source, as an approximate visualization of the \Sunburstarc\ galaxy in the source plane. The conceptual reconstruction was produced by ray-tracing the position of each of the emission clumps from the image plane to the source plane through the lens model. The approximate location of the features was determined by taking into account the ray-tracing of the different images of the \Sunburstarc, taking advantage of the high magnification of some of the images from the northern arcs as well as the less-magnified but more complete images in the \west\ and \southeast. We then populate these approximate locations with circular artificial sources with colors and surface brightness approximately matching those of the composite color view in \autoref{fig:arc1}. The source image of the extended galaxy (Source~50) and connecting ``bridge'' to the north is a direct ray-tracing of the image plane color rendition. 

We identify 19 discrete clumps in the main source galaxy. 
Clump 1 (the LCE) is positioned in the source plane such that it is multiply-imaged 12 times (a 13th undetected image is predicted by the lens model to be a central demagnified image). 
The clumps occupy a region sized $\sim1\times2$~kpc in the source plane, with the LCE clump positioned at one edge of the galaxy.

The closest clump to the LCE, clump 5, is resolved from the LCE mainly in the \north\ arc; other arcs lack the resolving power in the direction connecting clumps~1--5. As \autoref{fig:confusogram} showcases, the eastern parts of the source are not imaged in the \north\ arc. Clumps~1, 2, and 5--8 appear in all the images of the source except Images 5 and 6. Images 3 and 4 contain more of the source galaxy, including clumps~1--12, but lack the clumps at the east edge, which are seen in Images~7 and 8. Images 5 and 6 include only clump~1: Image~5 because it is only isolating clump~1 within its caustics (red caustics in \autoref{fig:caustics}); Image~6 would have included any emission west of clump~1 (i.e., west of the orange caustic in \autoref{fig:caustics}), if there was any. It does show an image of clump~101 which may be related to the same galaxy.
From color considerations and the source plane analysis, we identify that the non-leaking clump in Images 9 and 10 is clump~4. A lensing caustic can separate clumps~1 and 4 from the rest of the galaxy, thus creating their multiple images at these locations, as can be seen in \autoref{fig:caustics}. We identify an apparent double image of clump~4 in Image~8, which along with the discrepant clump (see Section~\ref{sec:weirdclump}) can be a result of a more complex caustic or a more complex source that is extremely magnified and tangentially resolved into several components only in this location. 
Finally, all the clumps appear in Images~11 and 12, which are complete images of the galaxy. However, due to the lower magnification, only the brightest clumps are readily identified. From their source plane positions, it is possible that the clumps we label 16 and 17 are the same as clump~11 or 12, but are less resolved, or that they are outside of the caustic of Images~3 and 4. 

For studies of the source properties of the \Sunburstarc\ galaxy, and in particular the physical properties of its LyC leaking versus non leaking regions, it is informative to know the source plane separation between each clump and the LCE. 
In \autoref{tab:clumpdist} we summarise this information. For each clump, we measure the source plane distance between the center of the clump and the LCE, and note in which of the twelve \Sunburstarc\ images it appears. The lower and upper limits on the distance to the LCE take into account the statistical lens modeling uncertainties from the MCMC, as well as the variations of source plane mapping from different images. The quoted distance is the average of the distances calculated from the $n$ different images of the same clump, i.e., $<d>_j = (1/n) \sum _{i=1}^n  d_{j,1}(i)$ where $d_{j,1}(i)$ is the distance between clump~$j$ and clump~$1$ in the source plane projection of Image~$i$.  The projections from Image~7 were excluded from this calculation due to the effect of modeling the foreground galaxy in this region on the lensing magnification. Including the source predictions from Image~7 would increase the upper uncertainty by a factor of $\sim2$. 

\vspace{-5pt} 
\subsubsection{Clump Sizes}
The extreme shear and anisotropic magnification, combined with the telescope PSF, complicates a traditional ray-tracing based source reconstruction of barely-resolved star forming regions. Simply ray-tracing unresolved lensed image to the source plane can only place upper limits on the clump size, as it will return the telescope PSF corrected for the linear magnification. 
A properly unmagnified, undistorted source plane can be recovered by using forward modeling techniques \citep[e.g.,][]{johnson17}, where a parameterized model of the source is constructed, ray-traced through the lens equation to the observer, the raw lensed image is convolved with the telescope PSF, and perturbed by noise and other instrument effects. The resulting image is then compared to the observed data iteratively to solve for the set of source plane model parameters that best minimizes the scatter between the model and the data. These and other techniques have been demonstrated to resolve structures down to tens of parsecs, especially when the magnification is particularly high \citep[e.g.,][]{zitrin2011,johnson17,vanzella2017,welch22}. 
A full forward-modeling of the \Sunburstarc\ is saved for a future publication. 
We can however deduce an upper limit for clump sizes from their apparent circular morphology and size in the image plane.

The different directions in which the source is distorted in the different images of the same part of the \Sunburstarc\ source galaxy allow for not only a tomographic analysis of the galaxy morphology, but also for a tomographic measurement of clump sizes, i.e., to constrain the spatial extent of individual star forming regions in multiple directions. We show this for each of the 19 clumps we identify, in \autoref{fig:psf}. 
For each of the images of each clump, we place a circle with $R=0\farcs08$ in the image plane, representing the \hst/ACS PSF. We ray-trace the PSF circle to the $z=\zarc$ source plane using the lens equation and the lens model outputs. For clarity, we only plot the results from the best fit model.  With the exception of image 1.5 of the LCE, plotting the statistical uncertainties would have resulted in only slight broadening of the shown ellipse lines. We find that typically, the \hst\ PSF circles de-lens into high-eccentricity ellipses in the source plane, with semi-minor axes of order 
$<50$~pc, 
and semi-major axes of order 
$500$~pc.
This means that to be unresolved or barely resolved in the image plane, a clump must be smaller than 
$50$~pc 
in the source plane. 
Regions with larger spatial extent in the source plane should appear extended and elongated in the image plane, as they span several source plane resolution elements. We note that the distinct star forming clumps observed in the \Sunburstarc\ are largely barely resolved.

Our upper limits on star forming clump sizes are significantly smaller than those reported for unlensed galaxies at cosmic noon, which is consistent with expectations from numerical simulations. For example, \citet{Meng20} conclude that clump sizes measured in field galaxies are overestimated by a factor of $2-3$ due to combination of the insufficient resolution of the telescope (usually \hst) and projection effects. They show that without the benefit of high lensing magnification, the observed clumps are likely an aggregate of several smaller clumps. 

Since the LCE clump is of high interest in the literature, we plot the delensed PSF for it separately, in \autoref{fig:psf_LCE}, and include lines from the 100 models that sample the parameter space. The projection from each image of the clump is shown in a different shade of gray; 
image 1.5 is not shown, because the high magnification uncertainty due to a foreground galaxy prevents it from providing useful constraints.
The clump, which is unresolved in our \hst\ imaging data, is constrained to 
$r\lesssim20$~pc 
in its short axis and 
$r\lesssim50$~pc 
in its long axis, leading to an effective radius of 
$r_{\rm eff}\lesssim\sqrt{20\times50}=32$~pc.
The estimate of \citet{vanzella20a} of $r_{e}\lesssim20$ in the tangentially-resolved direction, which they obtained by measuring the image plane size and assuming $\mu>25$, is consistent with our lens model-based upper limit.

\subsubsection{Evidence for galaxy interaction in the plane of the \Sunburstarc\ }

A likely companion to the \Sunburstarc, source~50 appears at the same redshift and $\sim 6$~kpc in projection north of the LCE clump. In velocity space, source 50 and the \Sunburstarc\ galaxy are separated by $50 \pm 28$~km~s$^{-1}$, indicating that they are likely interacting. Unlike the \Sunburstarc, this galaxy does not exhibit strong emission lines indicative of ongoing star formation; while a clear color gradient within this galaxy is apparent, overall it is redder and significantly less clumpy than the \Sunburstarc. 

As can be seen in the source plane projection in \autoref{fig:source}, the \Sunburstarc\ and its neighbor galaxy appear to be connected with a ``bridge'' of faint and largely diffuse emission. This emission could be a tidal tail resulting from a recent  interaction between these two galaxies. Another clump, labeled 101 in our figures and tables, is projected $\sim2$~kpc from the LCE but in a direction perpendicular to the line connecting the \Sunburstarc\ and galaxy 50. This clump has no direct spectroscopic redshift measurement, but based on the lensing analysis its redshift is consistent with that of the \Sunburstarc. It is therefore likely another correlated structure. 

This likely interaction between the \Sunburstarc\ and nearby galaxies could explain its clumpy nature. Analyses of zoom-in hydro-cosmological simulations indicate that minor galactic mergers can induce formation of clumps, and that massive clumps observed at large radii were likely formed ex-situ and merged \citep[e.g.,][]{Mandelker14,Mandelker17,Meng20}.

Other than the \MUSE\ data, the images of source 50 were not targeted for detailed spectroscopy (e.g., by our MagE or FIRE programs). Deep, high resolution IFU observations targeting both galaxies could inform an investigation of their interaction properties through spatially resolved velocity structure \citep[e.g.,][]{wuyts14}. 
The anticipated 6-filter, 1.4 h \JWST\ NIRCam imaging, which is planned for Cycle~1 (JWST-GO program\,02555, PI: Rivera-Thorsen) will provide depth and high resolution, and, combined with the existing multi-band \hst\ data, will enable spatially-resolved SED fitting of the two galaxies and the bridging emission in restframe-UV through NIR (0.08 -- 1.3$\mu m$). Such SED analysis will map these galaxies' metallicity, dust, star formation rate and history, and determine whether a recent burst of star formation occurred, as might be triggered by an interaction. It may provide clues to whether galaxy interactions play a role in LyC escape from star forming galaxies such as the \Sunburstarc.

\begin{deluxetable}{lllllll} 
\tablecolumns{7} 
\tablecaption{Source Plane Distances Between Clumps and the LCE \label{tab:clumpdist}} 
\tablehead{\colhead{ID} &
            \colhead{Distance from }    & 
            \colhead{Images}       &
            \colhead{Notes}             \\[-8pt]
            \colhead{} &
            \colhead{Clump~1 [kpc]}     &            
            \colhead{}             &
            \colhead{}             }
\startdata 
\input{clumpdist}
\enddata 
\tablecomments{Average projected distance in kpc between each of the non-LyC leaking clumps and the LCE (clump~1) in the source plane of the \Sunburstarc\ galaxy. The quoted distance is measured as the average of the source plane distances predictions from different multiple images. Uncertainties take into account both the image-to-image variation and the strong lens modeling uncertainty. }
\end{deluxetable}

\section{Summary} \label{sec:summary}
We present new multi-band \HSTlong\ imaging data and a lensing analysis of \clustername, a $z=\zcluster$ cluster lensing the \Sunburstarc\ galaxy. In addition to the \Sunburstarc, we identify multiple images of 14 strongly lensed galaxies, and measure spectroscopic redshifts of four of them from archival \MUSElong\ on the \vlt. The main arc in this line of sight, the \Sunburstarc, is a lensed image of a star-forming, LyC leaking galaxy at $z=\zarc$, which is lensed into ten partial images and two complete images. In particular, we confirm the identification of 12 images of the LyC emitting clump.
We model the field using the public software \lenstool\ \citep{jullo07}, a parametric algorithm that uses MCMC sampling of the parameter space. 
The lensing analysis provides a measurement of the mass distribution of the foreground lensing cluster, and the magnification and distortion solutions for analyses of the lensed sources behind it. Our findings are summarised as follows. 
\begin{itemize}
\item The projected mass density of the lens cluster \clustername, enclosed within $R=250$~kpc, is $M(<250 {\rm kpc})= $ \clustermass. We find this detailed lens model-based measurement to be in good agreement with previous estimates, and in particular, with the mass inferred from the arc radius; this agreement implies a symmetric mass distribution. The \chandra\ X-ray data indicate that the cluster is relaxed, and has a cool core, consistent with the star-formation activity identified in the BCG from imaging and spectroscopy. 

\item We measure the average magnification affecting the complete images of the \Sunburstarc\ galaxy,  $<\!\!\mu_{\rm W}\!\!>=13.5 ^{+2.4}_{-1.0}$ and
$<\!\!\mu_{\rm SE}\!\!>=13.1 ^{+1.0}_{-0.4}$, for the \west\ and \southeast\ arcs, respectively. Uncertainties are measured from $\sim100$ models representing a $1-\sigma$ sampling of the parameter space.   

\item The source galaxy of the \Sunburstarc\ is clumpy, with 19 emission clumps identified and mapped between its different images. The magnifications acting upon individual clumps within the \Sunburstarc\ range from a factor of a few, to $\mu>100$ in regions close to the critical curve. We provide the lensing magnifications and statistical uncertainties of the images of the identified emission clumps, and the average distance between each clump and the LCE clump.
\item A discrepant unresolved clump in the \northwest\ arc of the \Sunburstarc\ shows significantly different colors and spectral features from other similarly bright point sources. Previous work claimed that it is a transient, since it does not appear in the other images of the galaxy \citep{vanzella2020b}. We conduct a time delay analysis of the field, and show that the relative time delays between images 1-11 of the source span only several months to a year, a time span covered by our observations. A counter image of the discrepant clump would have been detectable in our data had it been a transient event. However, such a source does not appear or disappear in any of the images of the \Sunburstarc.  Our analysis instead confirms the conclusion of \citet{diego22}, that the occurrence of this clump (dubbed ``\textit{Godzilla}'' in their publication) is due to extreme magnification at this location, not due to time variability.    
\item We show that the lensing potential of \clustername\ results in multi-direction distortion of the source galaxy, providing tomographic resolving power to constrain the morphology of the source galaxy of the \Sunburstarc. We show that the source is probed in three distinct axes. While a full computed tomography and forward modeling source reconstruction analyses are left for future work, we use the lensing information to generate a qualitative reconstruction of the source galaxy by hand. We show an ``artist's impression'' realization of the source galaxy, with the emission clumps painted onto the source plane with similar colors as the observed image and at approximate positions as determined by projecting the twelve images to the source plane. We show conceptually how the source galaxy is bisected by the source plane caustics, thus forming the twelve observed images. 
\item Based on the projected source plane size of the \hst\ PSF, combined with the multi-direction sampling of the source plane by the lensing distortion, we place an upper limit on the size of any unresolved clump of 
$r\lesssim50$~pc.  
In particular, the unresolved LCE clump is constrained to 
$r_{\rm eff}\lesssim32$~pc. 
That is, to our knowledge, the most precise localization of a LyC escape site in any galaxy, an order of magnitude better than other observed LyC leaking galaxies so far; it showcases the unique power of gravitational lensing in the study of the mechanisms enabling ionizing escape.

\item We identify a likely companion galaxy, located $\sim 6$~kpc north of the \Sunburstarc\ in the source plane, labeled source~50 and observed near the \west\ and \southeast\ arcs.  In velocity space, source~50 is separated from the \Sunburstarc\ galaxy by 36~km~s$^{-1}$, indicating that the galaxies could be interacting. Future IFU observations of the two galaxies can map their velocity structures, while a combination of the existing \HST\  and the future \JWST\ multi-band imaging will enable spectral energy distribution fits to measure spatially resolved stellar populations, ages, metallicities, and other diagnostics, and examine the role of galaxy-galaxy interactions in prompting the escape of LyC radiation.
\end{itemize}

The multi-band \hst\ imaging, combined with spectroscopy from ground-based and space based observatories, enable the detailed analysis of the lensing potential of \clustername\ presented here, which in turn facilitates its use as a powerful cosmic telescope to study the background Universe.
Our model is and will be utilized by several ongoing and future works.
The primary lensed galaxy in this field, the extremely bright \Sunburstarc, is a topic of numerous publications, primarily because it is the most spatially-resolved example of LyC escape at any redshift. With an Einstein radius of $\sim30''$ and complex critical curve, this line of sight is a treasure trove for studies of intermediate and high redshift lensed galaxies and cluster physics alike.


\vspace{20pt} 
\begin{center}
ACKNOWLEDGMENTS
\end{center}
\noindent Based on observations made with the NASA/ESA {\it Hubble Space Telescope}, obtained at the Space Telescope Science  Institute, which is operated by the Association of Universities for Research in Astronomy, Inc., under NASA contract NAS 5-26555. These observations are associated with programs GO-15101, GO-15418, GO-15377, and GO-15949.
The authors acknowledge support from Programs GO-15101, GO-15949, GO-15337, provided through grants from the STScI under NASA contract NAS5-26555.
Based on VLT/MUSE observations collected at the European Southern Observatory under ESO programme 297.A-5012(A), PI: Aghanim, obtained from the ESO Science Archive Facility.
Support for this work was also provided by the National Aeronautics and Space Administration through Chandra Award Number GO8-19084X issued by the \Chandra\ X-ray Center, which is operated by the Smithsonian Astrophysical Observatory for and on behalf of the National Aeronautics Space Administration under contract NAS8-03060. The scientific results reported in this article are based on observations made by the \Chandra\ X-ray Observatory, and this research has made use of software provided by the \Chandra\ X-ray Center (CXC) in the application package, CIAO.
This paper used data gathered with the 6.5 m Magellan Telescopes located at Las Campanas Observatory, Chile. We thank the staff of Las Campanas for their dedicated service, which has made possible these observations. 
We thank the anonymous referee for a thoughtful and constructive review of the manuscript.
\vspace{20pt}

\facilities{HST(ACS,WFC3)}


\software{Drizzlepac\footnote{\url{http://www.stsci.edu/scientific- community/software/drizzlepac.html}}, Source Extractor \citep{sextractor}, \lenstool\ \citep{jullo07}, MAAT \citep{Ofek2014}, CIAO \citep[v4.13;][]{Fruscione2006}         }

\bibliographystyle{yahapj}
\bibliography{bib}

\end{document}

%% file: planckarc_arc1_mag.tex
1.1 & 237.530818 & -78.182515 &  34.7 & [29 -- 38] \\
1.2 & 237.525561 & -78.182754 &  82.7 & [75 -- 97] \\
1.3 & 237.524859 & -78.182817 &  58.7 & [56 -- 71] \\
1.4 & 237.519097 & -78.183181 &  15.3 & [9.3 -- 23] \\
1.5 & 237.518255 & -78.183256 &  2.3 & [0.5 -- 10.0] \\
1.6 & 237.517455 & -78.183354 &  13.2 & [7.6 -- 23] \\
1.7 & 237.509198 & -78.184647 &  4.3 & [3.9 -- 5.0] \\
1.8 & 237.501508 & -78.186259 &  29.6 & [27 -- 34] \\
1.9 & 237.499785 & -78.186745 &  27.3 & [25 -- 34](*) \\
1.10 & 237.498917 & -78.187077 &  30.1 & [28 -- 37](*) \\
1.11 & 237.493859 & -78.190767 &  12.1 & [12 -- 13] \\
1.12 & 237.562830 & -78.196471 &  13.4 & [12 -- 16] \\
2.1 & 237.528078 & -78.182604 &  128.9 & [92 -- 128] \\
2.2 & 237.527379 & -78.182635 &  109.4 & [127 -- 180] \\
2.3 & 237.523906 & -78.182902 &  15.0 & [13 -- 16] \\
2.4 & 237.520126 & -78.183100 &  12.9 & [11 -- 14] \\
2.7 & 237.508784 & -78.184748 &  6.0 & [4.8 -- 6.3] \\
2.8 & 237.503795 & -78.185731 &  42.8 & [35 -- 49] \\
2.11 & 237.493953 & -78.190812 &  12.0 & [12 -- 13] \\
2.12 & 237.563214 & -78.196386 &  13.2 & [12 -- 15] \\
3.3 & 237.523472 & -78.182926 &  13.0 & [11 -- 14] \\
3.4 & 237.520701 & -78.183058 &  15.5 & [13 -- 17] \\
3.7 & 237.508271 & -78.184839 &  12.5 & [9.2 -- 13] \\
3.8 & 237.504896 & -78.185501 &  51.2 & [40 -- 59] \\
3.11 & 237.493989 & -78.190867 &  12.3 & [12 -- 13] \\
3.12 & 237.563459 & -78.196326 &  13.1 & [12 -- 15] \\
4.3 & 237.523614 & -78.182962 &  15.5 & [14 -- 17] \\
4.4 & 237.520679 & -78.183110 &  15.7 & [13 -- 17] \\
4.7 & 237.509090 & -78.184749 &  5.2 & [4.3 -- 5.7] \\
4.8 & 237.502257 & -78.186149 &  35.5 & [32 -- 41] \\
4.9 & 237.499970 & -78.186785 &  31.1 & [28 -- 40](*) \\
4.10 & 237.499155 & -78.187089 &  33.9 & [31 -- 44](*) \\
4.11 & 237.494057 & -78.190672 &  11.1 & [11 -- 12] \\
4.12 & 237.563109 & -78.196465 &  12.9 & [12 -- 15] \\
5.1 & 237.530472 & -78.182525 &  33.7 & [28 -- 35] \\
5.2 & 237.525924 & -78.182728 &  53.0 & [48 -- 62] \\
5.3 & 237.524665 & -78.182835 &  35.8 & [33 -- 41] \\
5.4 & 237.519300 & -78.183166 &  13.3 & [8.6 -- 17] \\
6.1 & 237.530282 & -78.182473 &  36.5 & [31 -- 38] \\
6.2 & 237.526374 & -78.182639 &  34.4 & [31 -- 39] \\
7.1 & 237.529282 & -78.182482 &  32.1 & [27 -- 37] \\
7.2 & 237.527034 & -78.182569 &  43.0 & [40 -- 50] \\
7.3 & 237.523693 & -78.182846 &  10.9 & [9.7 -- 12] \\
7.4 & 237.520049 & -78.183013 &  12.2 & [10 -- 13] \\
7.7 & 237.507458 & -78.184894 &  45.5 & [26 -- 47] \\
7.8 & 237.505953 & -78.185193 &  59.6 & [52 -- 66] \\
7.11 & 237.493967 & -78.191389 &  16.3 & [16 -- 18] \\
7.12 & 237.563746 & -78.196152 &  13.9 & [13 -- 16] \\
8.1 & 237.528607 & -78.182557 &  48.8 & [38 -- 53] \\
8.2 & 237.527136 & -78.182623 &  62.2 & [61 -- 79] \\
8.3 & 237.523917 & -78.182875 &  14.1 & [13 -- 16] \\
8.4 & 237.520032 & -78.183080 &  12.6 & [10 -- 14] \\
8.7 & 237.508469 & -78.184775 &  8.4 & [6.4 -- 8.5] \\
8.8 & 237.504388 & -78.185573 &  44.7 & [36 -- 51] \\
8.11 & 237.493900 & -78.190958 &  13.3 & [13 -- 14] \\
8.12 & 237.563346 & -78.196316 &  13.4 & [12 -- 15] \\
9.3 & 237.523676 & -78.182928 &  14.1 & [12 -- 15] \\
9.4 & 237.520492 & -78.183090 &  14.3 & [12 -- 16] \\
9.8 & 237.504180 & -78.185652 &  45.7 & [37 -- 53] \\
10.3 & 237.523251 & -78.182942 &  13.4 & [12 -- 14] \\
10.4 & 237.521002 & -78.183046 &  18.7 & [16 -- 21] \\
10.7 & 237.508163 & -78.184876 &  15.7 & [11 -- 16] \\
10.8 & 237.505064 & -78.185473 &  53.6 & [42 -- 61] \\
11.3 & 237.523067 & -78.182913 &  11.7 & [10 -- 13] \\
11.4 & 237.521062 & -78.182998 &  19.4 & [17 -- 21] \\
12.3 & 237.522753 & -78.182948 &  16.0 & [13 -- 17] \\
12.4 & 237.521510 & -78.182999 &  37.0 & [32 -- 48] \\
13.7 & 237.507614 & -78.185015 &  45.9 & [32 -- 46] \\
13.8 & 237.506255 & -78.185277 &  101.1 & [82 -- 123] \\
14.7 & 237.507415 & -78.185061 &  74.4 & [51 -- 75] \\
14.8 & 237.506400 & -78.185256 &  123.9 & [102 -- 161] \\
15.7 & 237.507203 & -78.185102 &  149.0 & [97 -- 148] \\
15.8 & 237.506595 & -78.185218 &  180.0 & [159 -- 277] \\
16.11 & 237.494050 & -78.191144 &  14.6 & [14 -- 16] \\
16.12 & 237.563858 & -78.196174 &  13.5 & [12 -- 15] \\
17.11 & 237.494158 & -78.191336 &  15.6 & [15 -- 17] \\
17.12 & 237.564375 & -78.196021 &  13.7 & [12 -- 15] \\
18.11 & 237.494696 & -78.192019 &  15.9 & [15 -- 17] \\
18.12 & 237.565800 & -78.195482 &  16.3 & [14 -- 18] \\
19.11 & 237.494679 & -78.192066 &  16.1 & [15 -- 18] \\
19.12 & 237.566061 & -78.195415 &  16.6 & [15 -- 18] \\
\hline
51.11$^1$ & 237.495214 & -78.192641 &  13.5 & [13 -- 15] \\
51.12 & 237.567111 & -78.194878 &  22.8 & [19 -- 25] \\
52.11 & 237.495142 & -78.192566 &  12.3 & [12 -- 14] \\
52.12 & 237.566914 & -78.194972 &  21.4 & [18 -- 24] \\
53.1 & 237.570080 & -78.193511 &  1246.8 & [83 -- ...] \\
53.2 & 237.572192 & -78.191494 &  726.5 & [29 -- ...] \\
53.3 & 237.572516 & -78.191086 &  32.9 & [24 -- 37] \\
\hline
101.1$^2$ & 237.532701 & -78.182363 &  37.8 & [27 -- 49] \\ 
101.6 & 237.517073 & -78.183258 &  17.7 & [11 -- 36] \\ 
101.8 & 237.503649 & -78.185600 &  34.8 & [29 -- 39] \\ 
101.11 & 237.493719 & -78.191473 &  17.2 & [16 -- 19] \\ 
101.12 & 237.563151 & -78.196242 &  14.5 & [13 -- 17] \\ 

%% file: planckarc_arcs.tex
21.1 & 237.507879 & -78.195669 & \nodata  & $2.00_{-0.03}^{+0.03}$ & \\ 
21.2 & 237.563285 & -78.184445 &  & & \\ 
22.1 & 237.507377 & -78.195596 &  & & \\ 
22.2 & 237.563645 & -78.184605 &  & & \\ 
\hline
31.1 & 237.501071 & -78.194245 & 2.4600  & \nodata &\\ 
31.2 & 237.572190 & -78.186764 &  & & \\ 
32.1 & 237.501173 & -78.194403 &  & & \\ 
32.2 & 237.570735 & -78.186045 &  & & \\ 
\hline
41.1 & 237.508731 & -78.189852 & 1.1860  & \nodata&\\ 
41.2 & 237.558868 & -78.193976 &   & &\\ 
\hline
51.11 & 237.495214 & -78.192641 & 2.3709 & \nodata&Companion galaxy to the \Sunburstarc \\ 
51.12 & 237.567111 & -78.194878 &  &  &\\ 
52.11 & 237.495142 & -78.192566 &  &  &\\ 
52.12 & 237.566914 & -78.194972 &  &  &\\ 
53.1 & 237.570080 & -78.193511 &  & &\\ 
53.2 & 237.572192 & -78.191494 &  &  &\\ 
53.3 & 237.572516 & -78.191086 &  &  &\\ 
\hline
c61.1 & 237.538392 & -78.190088 &  \nodata & \nodata &Radial arc, Candidate \\ 
c61.2 & 237.545362 & -78.188855 &  & &\\ 
c61.3 & 237.493848 & -78.194733 &  &  &\\ 
c61.3 & 237.485901 & -78.196788 &  &  &\\ 
\hline
71.1 & 237.544744 & -78.190197 & \nodata & $2.20_{-0.06}^{+0.04}$ & Radial arc\\          
71.2 & 237.541128 & -78.190545 &  &  &\\ 
72.1 & 237.545902 & -78.189977 &  &  &\\ 
72.2 & 237.539452 & -78.190676 &  &  &\\ 
72.3 & 237.480527 & -78.192377 &  &  &\\ 
\hline
81.1 & 237.571984 & -78.194627 & \nodata & $2.14_{-0.01}^{+0.05}$ & \\ 
81.2 & 237.501340 & -78.190593 &  &  &\\ 
82.1 & 237.571210 & -78.195048 &  &  &\\ 
82.2 & 237.501364 & -78.190318 &  &  &\\ 
83.1 & 237.571123 & -78.195091 &  &  &\\ 
83.2 & 237.501366 & -78.190278 &  &  &\\ 
\hline
91.1 & 237.496776 & -78.193334 & 3.5100 &\nodata & \\ 
91.2 & 237.578283 & -78.188071 &  &  &\\ 
92.1 & 237.496803 & -78.193415 &  &  &\\ 
92.2 & 237.578121 & -78.187978 &  &  &\\ 
93.1 & 237.496902 & -78.193517 &  &  &\\ 
93.2 & 237.577906 & -78.187816 &  &  &\\ 
94.1 & 237.497320 & -78.193754 &  &  &\\ 
94.2 & 237.577150 & -78.187272 &  &  &\\ 
95.1 & 237.498103 & -78.194186 &  &  &\\ 
95.2 & 237.573889 & -78.185414 &  &  &\\ 
96.1 & 237.497953 & -78.194128 &  &  &\\ 
96.2 & 237.574173 & -78.185593 &  &  &\\ 
97.1 & 237.498285 & -78.194236 &  &  &\\ 
97.2 & 237.573730 & -78.185329 &  &  &\\ 
\hline
131.1 & 237.553434 & -78.191877 & \nodata &$1.52_{-0.02}^{+0.02}$ & \\ 
131.2 & 237.493765 & -78.187681 &  & & \\ 
132.2 & 237.492902 & -78.187827 &  & & \\ 
132.1 & 237.552558 & -78.191747 &  & & \\ 
\hline
141.1 & 237.537716 & -78.196427 & \nodata  & $1.76_{-0.04}^{+0.02}$ &\\ 
141.2 & 237.510804 & -78.182610 &  &  &\\ 
\hline
151.1 & 237.533169 & -78.197672 & \nodata  & $3.27_{-0.14}^{+0.14}$ &\\ 
151.2 & 237.518525 & -78.179783 &  &  &\\ 
\hline
161.1 & 237.531076 & -78.197533 & \nodata & $2.48_{-0.10}^{+0.03}$ &\\ 
161.2 & 237.517617 & -78.180774 &  &  &\\ 
162.1 & 237.530599 & -78.197467 &  &  &\\ 
162.2 & 237.518346 & -78.180675 &  &  &\\ 
\hline
c170.1 & 237.533765 & -78.182837 & \nodata  &\nodata & Candidate \\ 
c170.2 & 237.553368 & -78.198716 &  &  &\\ 
c170.3 & 237.523858 & -78.200091 &  &  &\\ 
\hline
181.1 & 237.556363 &  -78.180047 & \nodata &$3.15_{-0.08}^{+0.19}$  & Low confidence $z_{spec}$ 2.5820\\ 
181.2 & 237.518409 &  -78.196005 &  &  &\\ 
\hline
191.1 & 237.547077 &  -78.180167 & \nodata & $2.08_{-0.01}^{+0.12}$ & \\ 
191.2 & 237.524546 &  -78.194008 &  &  & \\ 

%% file: clumpdist.tex
  1 &  \nodata & 1,2,3,4,5,6,7,8,9,10,11,12 & LCE \\ 
  2 &  $0.60_{-0.32}^{+0.87}$ & 1,2,3,4,7,8,11,12 &\\ 
  3 &  $0.91_{-0.39}^{+1.12}$ & 3,4,7,8,11,12 &\\ 
  4 &  $1.05_{-0.16}^{+0.72}$ & 3,4,7,8,9,10,11,12 &\\ 
  5 &  $0.11_{-0.06}^{+0.25}$ & 1,2,3,4 &\\ 
  6 &  $1.08_{-0.16}^{+0.21}$ & 1,2 &\\ 
  7 &  $1.77_{-0.42}^{+0.82}$ & 1,2,3,4,7,8,11,12 &\\ 
  8 &  $0.83_{-0.31}^{+0.83}$ & 1,2,3,4,7,8,11,12 &\\ 
  9 &  $0.91_{-0.36}^{+0.81}$ & 3,4,8 &\\ 
 10 &  $1.21_{-0.42}^{+1.00}$ & 3,4,7,8 &\\ 
 11 &  $1.76_{-0.42}^{+0.97}$ & 3,4 &\\ 
 12 &  $1.70_{-0.44}^{+0.98}$ & 3,4 &\\ 
 13 &  $1.40_{-0.09}^{+0.24}$ & 7,8 &\\ 
 14 &  $1.45_{-0.09}^{+0.25}$ & 7,8 &\\ 
 15 &  $1.44_{-0.09}^{+0.25}$ & 7,8 &\\ 
 16 &  $1.20_{-0.32}^{+0.18}$ & 11,12 &\\ 
 17 &  $1.74_{-0.44}^{+0.24}$ & 11,12 &\\ 
 18 &  $3.70_{-0.74}^{+0.41}$ & 11,12 & ``Bridge''\\ 
 19 &  $3.73_{-0.74}^{+0.38}$ & 11,12 & ``Bridge''\\ 
 51 &  $5.86_{-1.21}^{+0.84}$ & 11,12 & Companion \\ 
 52 &  $5.61_{-1.18}^{+0.82}$ & 11,12 & Companion \\ 
 53 &  $6.37_{-1.12}^{+1.66}$ & North of 12 & Companion \\ 
101 &  $2.16_{-0.32}^{+0.46}$ & 1,6,8,11,12 & \\ 